\documentclass[twocolumn]{Adiab-4}

\usepackage{amsmath}
\usepackage{multirow}
\shorttitle{He Binary Stars}
\shortauthors{Zhang et al.}
\graphicspath{{./}}

\begin{document}

\title{Adiabatic Mass Loss In Binary Stars. IV. Low and Intermediate Mass Helium Binary Stars}
\correspondingauthor{Hongwei Ge}
\email{gehw@ynao.ac.cn}

\author[0009-0001-3638-3133]{Lifu Zhang}
\affiliation{Yunnan Observatories, Chinese Academy of Sciences, Kunming 650216, China}
\affiliation{University of Chinese Academy of Sciences, Beijing 100049, People's Republic of China}

\author[0000-0002-6398-0195]{Hongwei Ge}
\affiliation{Yunnan Observatories, Chinese Academy of Sciences, Kunming 650216, China}
\affiliation{Key Laboratory for Structure and Evolution of Celestial Objects, Chinese Academy of Sciences, P.O. Box 110, Kunming 650216, People's Republic of China}
\affiliation{International Centre of Supernovae, Yunnan Key Laboratory, Kunming 650216, People's Republic of China}

\author[0000-0001-5284-8001]{Xuefei Chen}
\affiliation{Yunnan Observatories, Chinese Academy of Sciences, Kunming 650216, China}
\affiliation{University of Chinese Academy of Sciences, Beijing 100049, People's Republic of China}
\affiliation{Key Laboratory for Structure and Evolution of Celestial Objects, Chinese Academy of Sciences, P.O. Box 110, Kunming 650216, People's Republic of China}
\affiliation{International Centre of Supernovae, Yunnan Key Laboratory, Kunming 650216, People's Republic of China}

\author[0000-0001-9204-7778]{Zhanwen Han}
\affiliation{Yunnan Observatories, Chinese Academy of Sciences, Kunming 650216, China}
\affiliation{University of Chinese Academy of Sciences, Beijing 100049, People's Republic of China}
\affiliation{Key Laboratory for Structure and Evolution of Celestial Objects, Chinese Academy of Sciences, P.O. Box 110, Kunming 650216, People's Republic of China}
\affiliation{International Centre of Supernovae, Yunnan Key Laboratory, Kunming 650216, People's Republic of China}

\begin{abstract}

The unstable mass transfer situation in binary systems will asymptotically cause the adiabatic expansion of the donor star and finally lead to the common envelope phase. This process could happen in helium binary systems once the helium donor star fills its Roche-lobe. We have calculated the adiabatic mass loss model of naked helium stars with a mass range of 0.35\,$M_{\odot}$ to 10\,$M_{\odot}$, and every mass sequence evolved from the He-ZAMS to the cooling track of white dwarf or carbon ignition. In consideration of the influence of stellar wind, massive helium stars are not considered in this paper. Comparing stellar radius with the evolution of the Roche-lobe under the assumption of conservative mass transfer, we give the critical mass ratio $q_{\textrm{crit}}=M_{\textrm{He}}/M_{\textrm{accretor}}$ as the binary stability criteria of low and intermediate-mass helium binary stars. On He-MS, the result shows $1.0<q_{\textrm{crit}}<2.6$, which is more unstable than the classical result of polytropic model $q_{\textrm{crit}}=3$. After early He-HG, the $q_{\textrm{crit}}$ quickly increases even larger than 10 (more stable compared with widely used result $q_{\textrm{crit}}=4$), which is dominated by the expansion of radiative envelope. Our result could be useful for these quick mass transfer binary systems such as AM CVns, UCXBs, and helium novae, and it could guide the binary population synthesis for the formation of special objects such as SNe Ia and GW sources.

\end{abstract}

\keywords{Binary evolution(154) --- Helium-rich stars(715) --- Stellar evolution(1599) --- Common envelope evolution(2154)}


\section{Introduction}
\label{sec-intro}
The observation has shown that a large rate of stars exist in binary systems (e.g., \citealt{2012Sci...337..444S,2017ApJS..230...15M,2024PrPNP.13404083C}). If those systems are close enough, with the evolution of the primary stars, the primaries will expand and fill its Roche-lobe. Finally it will starts the Roche-lobe overflow (RLOF). The magnitude of binary mass transfer has been proved in the last few decades. Mass transfer in binary systems creates variable events that can hardly be explained in the single-star systems such as double neutron stars (NSs) and black holes (BHs), type Ia supernova (SN Ia), and gravitational wave (GW) sources (\citealt{2020RAA....20..161H,2023pbse.book.....T}).

Depending on whether the primary star could be limited near the range of the Roche lobe during RLOF, the binary mass transfer could be separated into the stable phase and common envelope (CE) phase (\citealt{1976IAUS...73...75P}). For the first phase, the donor star remains in the Roche lobe and it stays in thermal equilibrium during the mass transfer stage until it loses enough material and shrinks smaller than its Roche lobe. Secondary could have enough time (longer than thermal or nuclear timescale) to accrete mass to become Algol-like binaries (\citealt{1955ApJ...121...71C}) or create the accretion-disk cataclysmic variables and low-mass X-ray binaries (LMXBs) (\citealt{2005ARA&A..43....1G}).
However, if the donor star loses the stability and considerably exceeds the Roche-lobe radius, a violent mass loss of the donor star will come out (\citealt{1997A&A...327..620S}). It breaks the thermal equilibrium and triggers the adiabatic expansion. In this case, the stellar radius expands considerably greater than binary separation. Such unstable mass transfer leads to the CE phase. 
The stability in binary mass transfer is a conclusive issue for the outcome of binary evolution. If the mass transfer is stable, the accretor will have enough time to increase mass. If unstable, the donor star will expand further than binary separation and form the CE. During the CE phase, the reduction of binary orbital energy overcomes the binding energy and ejects most of the donor envelope in a short time. CE evolution is able to create a short-period binary system or cause binary merge during CE eject. 
The difference in binary evolution between stable and unstable mass transfer leads to very different fates, such as progenitors of SNe Ia, the merge of double white dwarfs, and the formation of short-period hot subdwarf stars. Because the timescale is very short for the CE phase, it is difficult to find these systems during unstable mass transfer. Though, there are some potential remnants of CE ejection or CE merge (\citealt{2006A&A...451..223T,2011A&A...528A.114T,2013Sci...339..433I}). In the meantime, the accretion of a secondary star can also lead to uncontrollable expansion and cause CE evolution. In this article, we focus on the stability criteria of the donor and ignore the influence of such an accreting mechanism.

In the 1990s, the polytropic model was developed to solve the stability criteria problem (\citealt{1987ApJ...318..794H,1997A&A...327..620S}). It simplifies the stellar structure and is widely used in binary population synthesis  (\citealt{2002MNRAS.329..897H,2014A&A...563A..83C}). Other studies are trying to discover the stability criteria of some specific binary systems. Compared with studies of the population of observed systems, more results regarding the stability of mass transfer have been obtained (e.g., \citealt{1999A&A...350..928T,2002ApJ...565.1107P,2003MNRAS.341..669H,2019MNRAS.482.4592V,2020A&A...641A.163V,2021ApJ...908..229L}).
However, most of them have not given us the results in the complete parameter space.
In our previous Papers \citep{PaperI, PaperII, PaperIII}, we introduced the model sequences of adiabatic mass loss from the main sequence (MS) star to the Red Giant Branch (RGB) star and Asymptotic Giant Branch (AGB) star. The result shows a great improvement for RGB and AGB stars. Our studies are adjusted for the stability prediction of classical studies (e.g., \citealt{1965AcA....15...89P,1969ASSL...13..237P,1987ApJ...318..794H}) and have given a more stable parameter space for Giant-Branch (GB) stars and massive stars.
Recently, the 1D simulation of the mass-transfer evolution by MSEA code \citep{2023A&A...669A..45T} has simulated the mass-transfer evolution in the mass range from $1\,M_{\odot}$ to $8\,M_{\odot}$ and successfully expands the criteria to giant branch stars. Their result and our work have formed a good mutual confirmation on $M=1.0\,M_\odot$ and $M=5.0\,M_\odot$ Hertzsprung gap (HG) stars.

One inescapable case is the helium binary system. It contains a helium dominated core or shell burning star and a secondary. The helium stars have a faster life than normal stars with the same mass and could also become donors during RLOF. 
The helium star is formed by the ejection of the hydrogen envelope of the progenitor due to binary mass transfer or fast stellar wind. Different from a helium white dwarf (He WD), the center of a helium star is capable of starting helium ignition. The formation channels of helium stars have been studied by using the binary population synthesis method \citep{2002MNRAS.336..449H} and found as the dominant contribution of UV-upturn of elliptical galaxies \citep{2007MNRAS.380.1098H}. Low-mass helium stars, especially for stellar mass $M<1.0\,M_\odot$, are very likely to be observed as B/O type hot subdwarf stars (sdB and sdO). For massive helium stars $M>10\,M_\odot$, it is believed to refer to some objects like Wolf-Rayet stars (WR). 

sdB stars lay on the ultra-horizontal branch of the Hertzsprung–Russell diagram (HRD), denser and dimmer than O/B type MS stars (\citealt{2009ARA&A..47..211H,2016PASP..128h2001H}). Their high temperature and luminosity suggest a helium-burning center instead of hydrogen. Different from pure helium composite stars, a large amount of sdB stars have features of hydrogen absorption lines in the spectrum, which is believed to be the symbol of a thin hydrogen envelope ($<10^{-2}\,M_\odot$) near the stellar surface. Besides, sdBs are commonly found in binary systems (\citealt{2001MNRAS.326.1391M,2011MNRAS.415.1381C}). Most possible companions in short-period systems (P$<$15\,d) are white dwarfs (WD), and some systems could even be as short as 1 hr \citep{2022A&A...666A.182S}, which makes those sdBs possible ultra-compact X-ray binaries (UCXB) and low-frequency gravity wave (GW) source. In addition, one-third of long-period sdB binaries are composed of MS stars \citep{2018MNRAS.473..693V}. With the evolution of helium stars, some of the close sdB binary systems could experience RLOF. 

The formation of sdBs has a tight relationship with binary mass transfer. During this stage, the radius of the primary star (e.g. HG, RGB, AGB) could expand over its Roche lobe and the stellar mass flew through the inner Lagrangian point ($L_1$). In the last two decades, studies in stellar evolution simulation and binary population synthesis have drawn a good picture to show that mass transfer in close binary systems is a compelling case for removing most of the hydrogen envelope (the systems are also called stripped helium stars) and leave a thin hydrogen shell behind (\citealt{2020RAA....20..161H}). Due to the low luminosity and strong gravity on the sdB surface, the hydrogen envelope of low-mass helium stars could hardly be ejected by the stellar wind during most of the time of star evolution. 

WR binary systems are much more massive ($M>10\,M_\odot$) and brighter than sdB. Strong and broad emission lines in their spectrum suggest that the stellar wind significantly influences its stellar evolution \citep{2006A&A...457.1015H}. Because of this, a WR star might eject its hydrogen envelope through stellar wind instead of RLOF. Despite all this, almost all WR stars we have observed are in binary systems. 
Oddly, helium stars are barely found in those intermediate masses ($2.0\,M_\odot<M<10\,M_\odot$). It is possibly because of rare quantity (compared to sdBs) and less brightness (compared to WRs). The detection of HD 45166, believed to be a "quasi Wolf-Rayet" (qWR) binary system \citep{2005A&A...444..895S}, first gives us a glance at these stars. Recently, medium-resolution spectra observation in Magellanic Clouds gives us 25 candidates of intermediate helium binary star with an MS companion \citep{2023ApJ...959..125G}. Their observation shows that intermediate-mass helium stars are similar to WR stars on the spectrum, but the evolution characteristic is close to low mass helium stars by the weak wind on He-MS. 

This article will use the adiabatic mass loss model to simulate the unstable mass transfer process of low and intermediate-mass helium stars. The massive helium star will be studied in the future. Then, we will analyze the stability criteria of helium binary systems by comparing the radius between the donor star and its Roche lobe. 
Section \ref{sec:sequences} will introduce the method of creating helium star sequences as donor stars. Based on different radius features, we divide helium stars into different stages and select several models for adiabatic mass loss. Section \ref{sec:adiab} gives the specific process of adiabatic mass loss of helium mass sequences. Firstly, we give a sample to introduce radius change during adiabatic mass loss on Radius-Mass diagram. Then we will give the critical mass ratio $q_{\textrm{crit}}$ for judging unstable mass transfer. Finally, Section \ref{summary} will be the summary of this article.

\section{build helium star sequences} \label{sec:sequences}

We must simulate variable helium stars in different masses and evolutionary stages to study adiabatic mass loss in the helium binary system. To develop a sequence of naked low and intermediate-mass helium stars, here we use \textit{STARS} code which was developed by \citet{1971MNRAS.151..351E,1972MNRAS.156..361E,1973MNRAS.163..279E} and \cite{2004PASP..116..699P}. It is a one-dimensional (spherically symmetric) non-Lagrangian code. We have already introduced this code in the second section of Paper I \citep{PaperI}.

\begin{figure*}[ht!]
\plotone{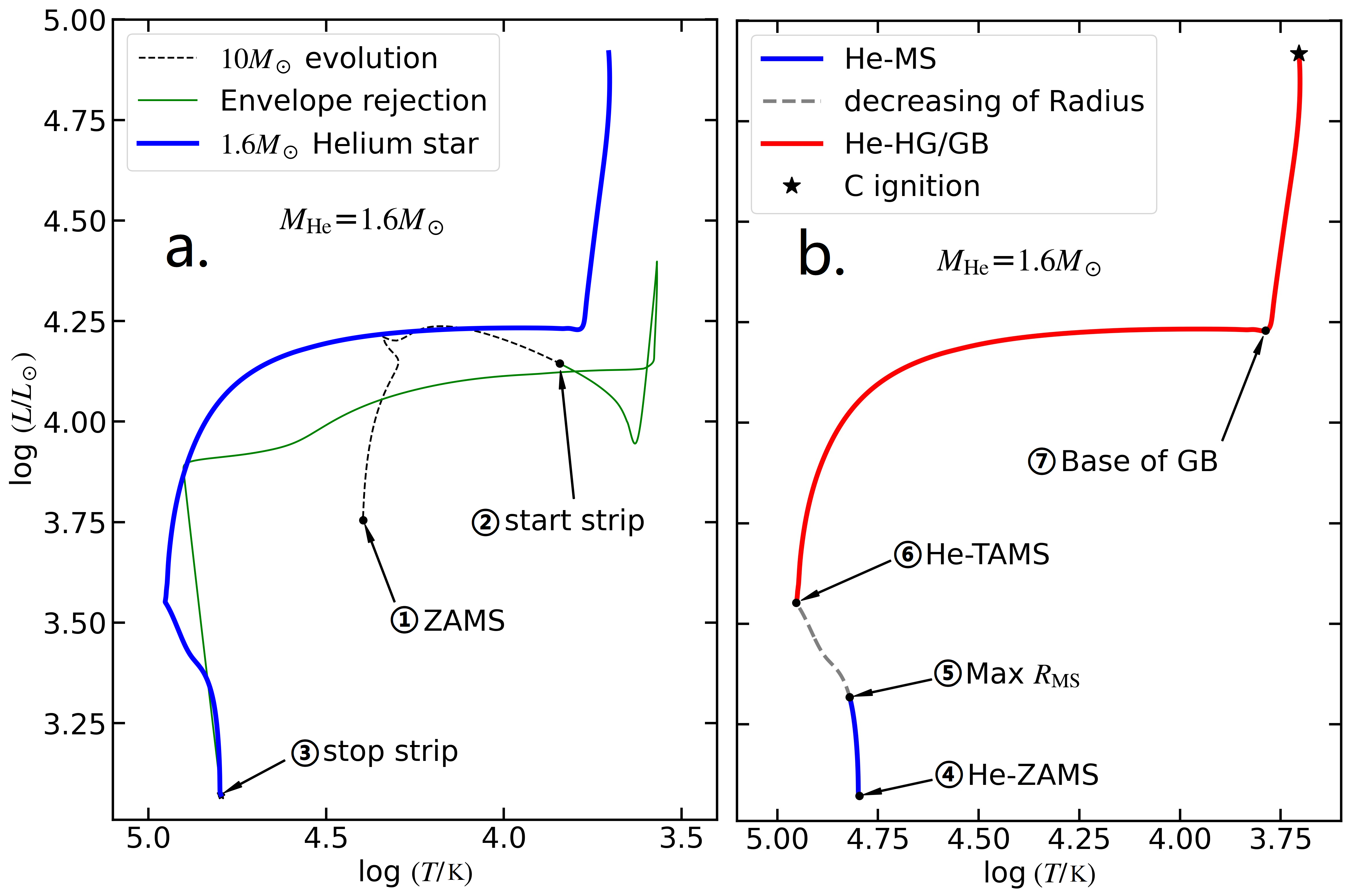}
\caption{Evolutionary track of a naked $1.6\,M_{\odot}$ helium star in the HRD. In this paper, 'log' represents base-10 logarithm.
In sub-figure 'a', we show the complete processes of building the helium star. The black dash line shows the $10\,M_{\odot}$ progenitor star from ZAMS to HG. Its hydrogen envelope is ejected during the green line. We stop the mass-losing stage until stellar mass reaches $1.6\,M_{\odot}$. The blue line shows the helium star evolution stage. The code is killed at carbon ignition. Here we set the carbon ignition limit at the of carbon luminosity when $L_{\rm C}>100\,L_{\odot}$. 
Sub-figure 'b' is the helium star evolution stage. It is divided into three phases by radius: Radius-increasing phase on He-MS, Radius-decreasing phase on He-MS, and He-HG/GB phase. 
\label{fig:1}}
\end{figure*}

In our simulation, we consider a normal Population I metallicity situation ($Z=0.02$). We use the overshooting parameter $\delta$=0.12 (\citealt{1997MNRAS.285..696S,1998MNRAS.298..525P}) and the mixing length parameter $\alpha$=2.0 \citep{1998MNRAS.298..525P}. They are good fits to the sun. In the case of focusing on low and intermediate-mass helium stars ($M<10\,M_{\odot}$), which more likely bring out weak wind around $10^{-6}\sim10^{-8}\,M_{\odot}\textrm{/yr}$ in most time of evolution, we override stellar wind during the evolution stage of helium burning. Since observation shows little evidence of helium star structure after the core burning period, no wind assumption brings another advantage: it makes more models evolve thicker shells at the later stage to help simulate larger stars to cover the unknown parameter space of helium binaries. On the other hand, helium stars have lower mass limits, just like normal low-mass stars. Once the total mass is lower than $0.35\,M_{\odot}$, the center of the star would not form the environment for helium ignition and finally lead to the He WD phase. To sum up, we settle the helium star mass sequences from $0.35\,M_{\odot}$ to $10\,M_{\odot}$.

The observation of sdB/O has ensured that a thin hydrogen shell widely exists on helium stars, though it is less than $10^{-3}\,M_{\odot}$ in most cases. Such a thin hydrogen shell cannot start a shell nuclear reaction and become a major influence of energy during the helium star evolution. In this article, we simply override the hydrogen shell and consider the naked helium star approximation to simplify the calculation. Producing and evolving a certain helium star experiences three parts in our simulation. Firstly, we build a typical zero-age main sequence (ZAMS) star and let it evolve through MS. After this part, a helium core is formed at the star's center. Secondly, at the HG stage, we stop abundance change due to helium and heavy metal ignition and start ejecting the hydrogen envelope by using multiple times of Reimers wind \citep{1975psae.book.....B}. It finally strips the envelope entirely and exposes the helium core inside. Then, by reducing and increasing the stellar mass, we can get the mass sequences of such naked helium stars. At last, once a certain helium core mass is gotten, stop wind mass loss and re-open the abundance change, such a helium star starts to evolve as a helium-zero-age main sequence (He-ZAMS) star.

\begin{figure}[ht!]
\plotone{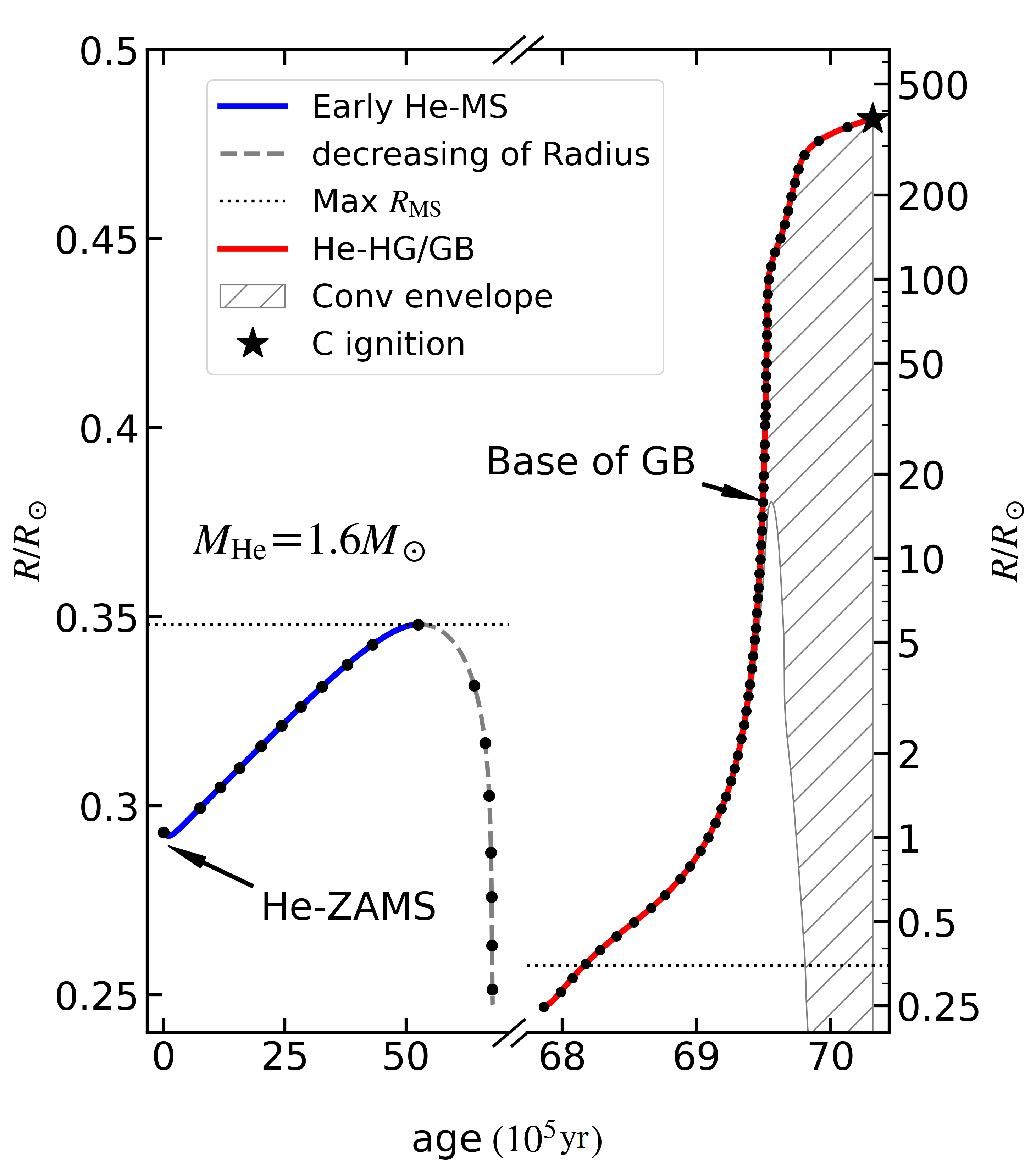}
\caption{Radius evolution of $1.6\,M_{\odot}$ helium star from He-ZAMS to C ignition. We change the ticks of He-MS and He-HG/GB to enhance the tendency of radius change. Dot lines represent the maximum radius of He-MS. The shadow area is the convective envelope zone. The models we choose for adiabatic mass loss are black dots.
\label{fig:1.6radius}}
\end{figure}

We take $1.6\,M_{\odot}$ helium star as a sample to introduce different stages of helium star evolution. Figure \ref{fig:1} shows the evolution track of a $1.6\,M_{\odot}$ helium star in the HRD. The method for creating the helium star in the previous introduction is shown in sub-figure 'a'. We focus on the helium star evolution stage particularly in sub-figure 'b'. 

The evolution of the stellar radius is shown in Figure \ref{fig:1.6radius}. The blue and red lines are the radius when the helium star is expanding, and the gray dash line is for shrinkage. The shadow area is the convective envelope zone.

After helium ignition at He-ZAMS, 3\,$\alpha$ reaction starts at the center, forming a convective core. Similar to the hydrogen burning process on the MS phase of intermediate-mass normal stars, helium burning finally burns off helium inside the convective zone and makes star maintain its luminosity. Such helium main sequence (He-MS) shall maintain for about a nuclear timescale until center helium is exhausted at the terminal age of helium main sequence (He-TAMS). 

On early He-MS, sustaining 3\,$\alpha$ reaction, the average relative atomic mass at the center convective core keeps increasing. It finally led to radius expansion. At the end of He-MS, the center helium fraction decreases lower than 0.2 so that 3\,$\alpha$ shall not be able to keep burning at the center. With a decrease of helium and increase of carbon in the center, the 4\,$\alpha$ reaction gradually takes advantage of later He-Ms and turns appreciable carbon into Oxygen. Conversely, star goes through the shrinkage stage at the later He-MS phase. At He-TAMS, the convective core gradually disappears, leaving a carbon/Oxygen core (CO core) behind. Helium star comprises a CO core and a helium envelope after He-TAMS.

After He-MS, the CO core star shrinks, and a helium-burning shell is developed at the bottom of the stellar envelope. At this phase, the helium star expands significantly like an ordinary star at HG. For those helium stars at a certain mass range, like $1.6\,M_{\odot}$ helium star, their expansion at this phase is so strong that developing a convective envelope similar to normal AGB stars. By the different structures of the helium envelope, we depart this phase into helium Hertzsprung Gap (He-HG) and helium Giant Branch (He-GB). The helium-burning region keeps supplying mass for the CO core during the He-HG and the He-GB phases.

On the one hand, more massive stars build up enough mass for carbon ignition. However, due to the roughness of elements and star nets, our work does not simulate the evolution after carbon ignition. Here, we technically set the limit of the carbon ignition when the luminosity of carbon ignition $L_{\rm C}>100\,L_{\odot}$. If the luminosity of carbon ignition is lower than this limit, the helium star will evolve to the shrinkage stage and end as a WD. If not, the fierce carbon ignition flame will cause a temperature spike inside of the core. It finally reaches the boundary of our program calculation. 

Carbon ignition is a complex process. For degenerate CO core cases, carbon ignition might be unstable and cause carbon flash events or become supernovae (SNe). In this paper, we roughly separate carbon flash and non-degenerate core (similar to more massive stars) by center degeneracy $\Psi_\mathrm{c}=3$ (\citealt{2017PASA...34...56D}). The carbon flash is similar to helium flash for some intermediate-mass RGB stars going through the horizontal branch. Simulation shows that carbon flash is off-center as well. It is still unclear whether the carbon flash will reverse the degenerate core like helium flash or lead to the electron-capture supernova (ECSN) phase. For helium star with a non-degenerate core, carbon ignition creates an onion structure inside the core, finally leading to the iron core-collapse supernova (CCSN). In the case of the very lifetime after carbon ignition, the helium envelope is unlikely to be wholly ejected. Finally it leaves helium emission lines in the supernova spectrum (\citealt{2023pbse.book.....T}). Both situations would quickly evolve to a supernova phase, which makes it worthless to consider binary mass transfer after carbon ignition. Therefore, our simulation stops at carbon ignition. On the other hand, a low-mass helium star could not start carbon ignition due to its thin envelope. When the flame of the burning shell reaches the surface of the helium star, helium ignition stops and forms a carbon/oxygen white dwarf (CO WD) in the end. However, we cannot give more details of the following evolution due to the small number of stellar shells and elements. The possible carbon flash and supernova phases are still directly unknown in our simulation.

\begin{figure*}[ht!]
\plotone{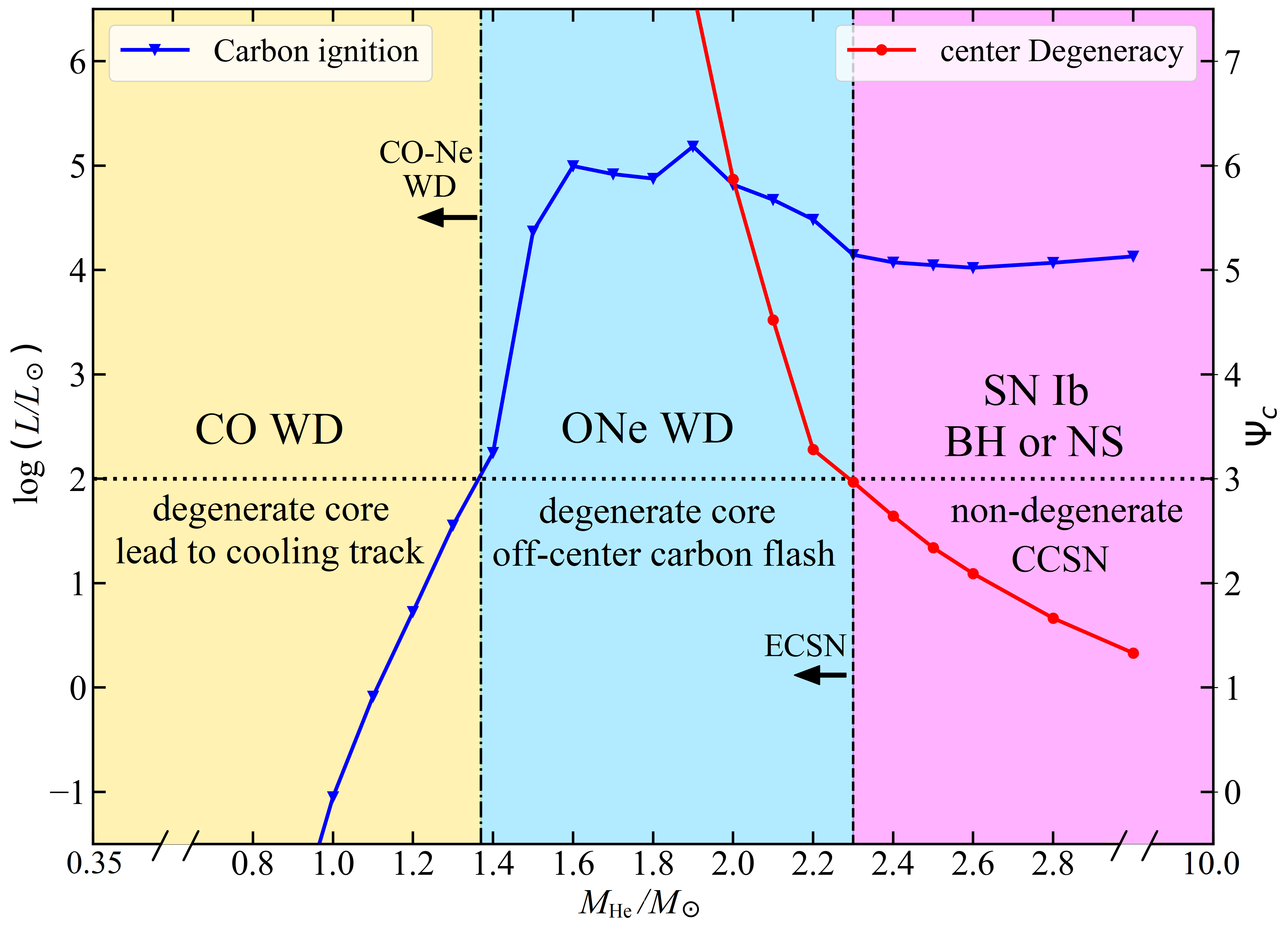}
\caption{The final status of helium stars. We roughly separate them by the carbon luminosity $L_{\rm C}$ and center degeneracy $\Psi_c$. For $M_{\textrm{He}}<1.4\,M_{\odot}$ stars, they will become CO WDs. For $1.4\,M_{\odot}<M_{\textrm{He}}<2.3\,M_{\odot}$ stars, off-center carbon flash reserves the degenerate core, and they end as ONe WDs. For $M_{\textrm{He}}>2.3\,M_{\odot}$ helium stars, carbon ignition starts in the center and they lead to CCSNe. Due to the helium envelope, the hydrogen lines do not exist in the spectrum (SN Ib). The special situations, CO-Ne WD and ECSN, can not be certain in our simulation. We roughly give their top limits in the figure. 
\label{fig:LZ}}
\end{figure*}

Here we compare our results with the final outcome of AGB stars by \citet{2017PASA...34...56D} and give a prediction of their ending. The final status of our helium star sequences is shown in Figure\,\ref{fig:LZ}.

For helium stars with $M_{\textrm{He}}<1.4\,M_{\odot}$, they will lead to the cooling track and end as CO WD. The carbon ignition power is always lower than $100\,L_{\odot}$. Such a weak power of carbon flash can not make the burning flame dredge into the center. It shall form a Ne shell inside the CO core, identified as a carbon/oxygen-neon white dwarf (CO-Ne WD).
For helium stars in the mass range of $1.4\,M_{\odot}<M_{\textrm{He}}<2.3\,M_{\odot}$, they have enough core mass for carbon ignition, and the flame of carbon flash is strong enough to dredge into the center of the degenerate core. After the carbon is exhausted, it finally turns into a degenerate oxygen/neon white dwarf (ONe WD). For some massive ONe WDs, if the degenerate core is heavier than around $1.375\,M_{\odot}$, the electron-capture process will happen and lead to ECSN. 
For more massive helium stars ($M_{\textrm{He}}>2.3\,M_{\odot}$), helium stars eventually form a non-degenerate CO core and start carbon and oxygen ignition in the center. It finally cause the iron-collapse event in the center and forms the CCSN. Due to the stars' stripped hydrogen envelope during the first mass transfer, the supernovae should not obtain hydrogen lines and become SNe Ib (\citealt{2023pbse.book.....T}). We also notice that the very detailed simulation by \citet{2019ApJ...878...49W} also gives the final outcomes for helium stars. The main difference is the consideration of the stellar wind. In our simulation, with out the wind, the minimum mass of carbon ignition will be lower due to the helium shell burning. Excluding that, the results are similar.

\begin{figure*}[ht!]
\plotone{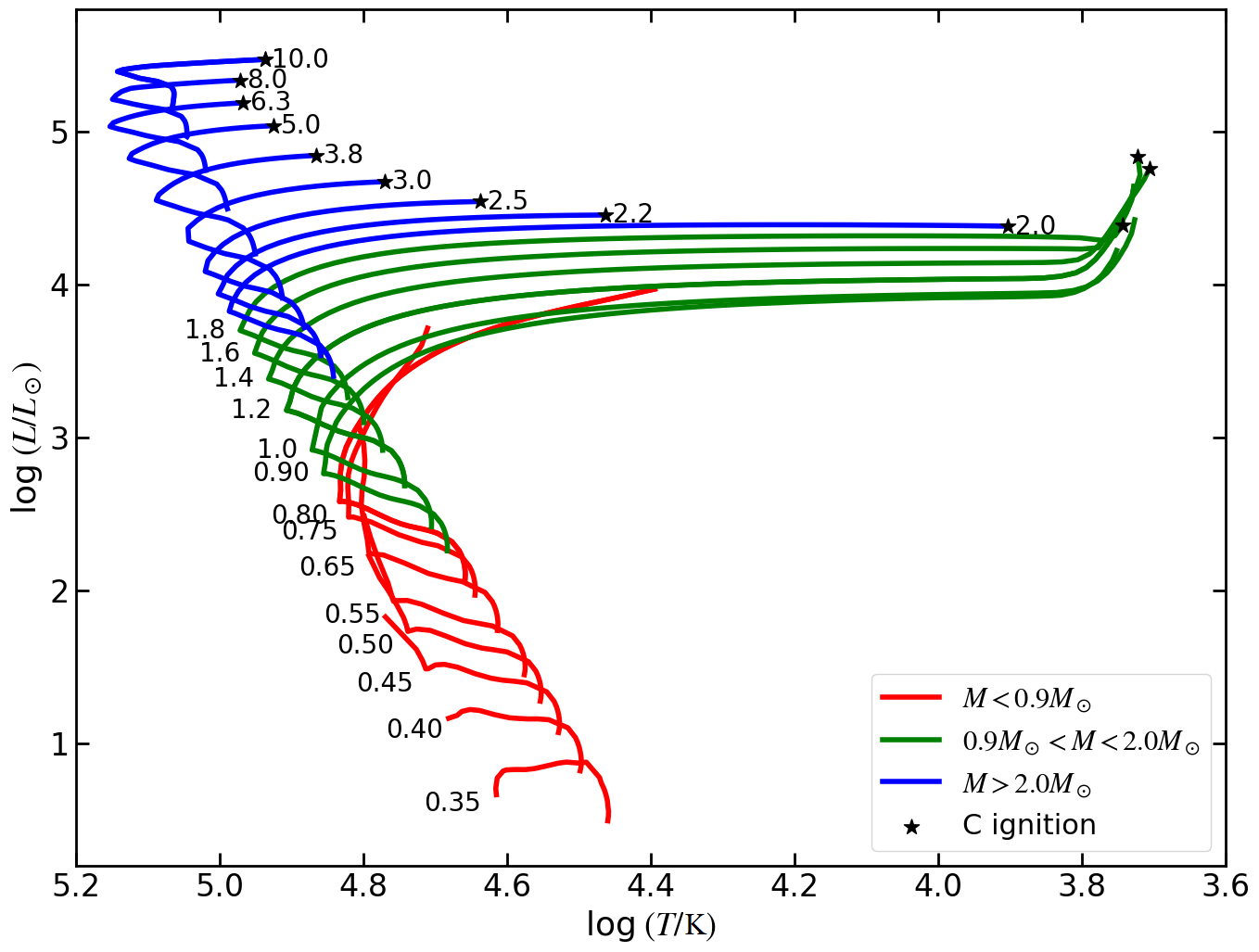}
\caption{Mass sequences of helium stars in the HRD. We divided them into three parts out of their envelope feature.
Blue lines: carbon ignition before developing convective envelope; 
Green lines: could evolve convective envelope; 
Red lines: envelope exhausted before developing convective envelope.
The models that start carbon ignition are shown as black stars.
\label{fig:2}}
\end{figure*}

To study the mass transfer stage, we pay more attention to the structures of the different helium stars. The main stages for mass transfer are limited to He-MS/He-HG/He-GB. Notably, some helium stars do not experience He-HG/He-GB.
Figure \ref{fig:2} shows some representative samples of evolution stages in our helium stars sequences. Different from the red giant branch (RGB) stars, only a narrow mass range ($0.9\,M_{\odot}<M_{\textrm{He}}<2.0\,M_{\odot}$) of helium star can evolve to He-GB phase. For $M_{\textrm{He}}>2.0\,M_{\odot}$ helium stars, they failed to develop a convective envelope before carbon ignition. 
For $M_{\textrm{He}}<0.9\,M_{\odot}$ helium stars, their helium envelopes are too thin to develop a convective zone before the cooling track of WD.
Another unique part lies on $M_{\textrm{He}}<0.6\,M_{\odot}$ helium stars: they barely expand during the He-HG phase out of the super-thin envelope. This radius behavior means their maximum radius on He-HG could be lower than it on He-MS.

To study mass transfer in close binary systems, we give more attention to the radius change of the donor star. According to the previous part, helium stars can divide into three stages based on radius change: the first expanding stage on early He-MS; the shrinkage stage on later He-MS and the second expanding stage on He-HG/GB. The symbols are shown in Figure\,\ref{fig:1}-b. To help adiabatic mass loss simulation, we pick multiple models on three stages of radius change and sign them up as bot markers in Figure\,\ref{fig:1.6radius}.

\section{Adiabatic Mass Loss} \label{sec:adiab}

We use the method in Paper I \citep{PaperI} to simulate the adiabatic mass loss situation for these selected models. In this part, we keep locking the entropy profile within the mass coordinate of every helium star and simulate stellar evolution during mass loss from surface to center. Such adiabatic mass loss could cause varying degrees of stellar expansion, and the stellar radius may finally be larger than the separation of the binary system. This process is likely to become the CE evolution phase.

We precisely simulate the adiabatic mass loss processes of different helium stars and give the stability criteria for the CE phase. Section \ref{subsec:1.6 adiab} gives a sample of $1.6\,M_{\odot}$ helium stars. We describe the method of getting stability criteria in this Subsection and finally sum up all mass sequences in Section \ref{subsec:qcrit}. In this section, we show our results as a critical mass ratio ($q_{\textrm{crit}}$) in the Mass-Radius diagram.

\subsection{Adiabatic Response And Stability Criteria} 
\label{subsec:1.6 adiab}
This section will show a sample of $1.6\,M_{\odot}$ helium star sequence. Like the description in Section \ref{sec:sequences}, $1.6\,M_{\odot}$ helium star could evolve through all stages from He-MS to He-GB. That makes it a representative case of study adiabatic mass loss at each stage. Our work is based on the adiabatic mass loss model, already introduced in Paper I \citep{PaperI}.

\begin{figure*}[ht!]
\plotone{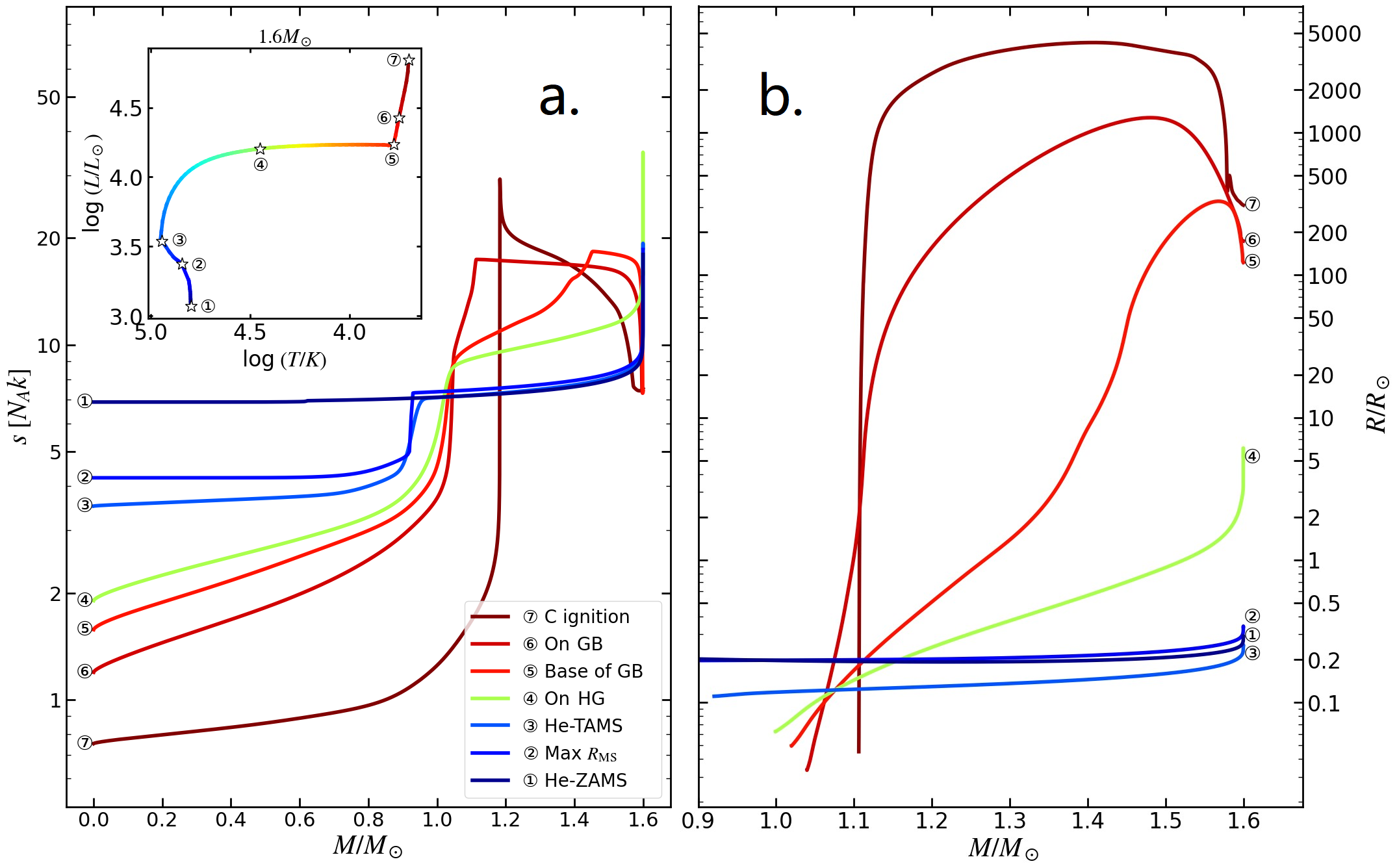}
\caption{In sub-figure 'a', we show the initial entropy profiles of $1.6\,M_{\odot}$ helium stars in different evolution stages. Sub-figure 'b' shows radius changes during adiabatic mass loss. 
\label{fig:1.6entropy-radius}}
\end{figure*}

To clear the most primary input quantities, we give the stellar entropy profiles in Figure \ref{fig:1.6entropy-radius}-a. It shows some representative evolution moment of $1.6\,M_{\odot}$ helium star. These moments are shown on the HRD in this figure. On the He-MS evolution stage, the convective core contracts with the decrease of the helium fraction, which causes a decrease in core entropy. Outside the core, entropy increases with the mass profile to show a much more stable structure than the convective zone. After He-MS, the convective core disappears due to the helium exhaustion at the center. Meanwhile, the core contract leads to gravitational potential energy release. At the He-HG stage, the helium envelope starts expanding and entropy increases rapidly by absorbing the gravitational potential energy from the center. Helium stars must keep a radiative envelope until the base of He-GB. Here, the surface area is getting too thin and cool to keep the structure. The convective envelope develops from surface to core and forms an equal-entropy zone inside this area. On the He-HG/GB stages, helium burning at the bottom of the envelope region enlarges the mass of the CO core. In this $1.6\,M_{\odot}$ case, the CO core shall keep electron degenerate. When the degenerate CO core reaches a certain status, carbon flash will start at a specific region (off-center). The final fate of different stars have been discussed in Section \ref{sec:sequences}. Before carbon ignition, the overly expanding convective envelope could significantly influence the efficiency of internal energy conduction. It finally leads to the formation of a negative (superadiabatic) specific entropy profile at this evolution stage.

We put these stellar models into adiabatic mass loss simulation, and the characteristic adiabatic response of a star to mass loss is shown in Figure \ref{fig:1.6entropy-radius}-b. In these cases, radius characteristics can be concluded into two modes by their envelope structure. Before the base of He-GB, the stellar envelope is radiative and keeps a relatively dense area near the surface. Entropy rapidly increases from the core to the surface. With the process of adiabatic mass loss, although it is experiencing adiabatic expansion, the cold and low-entropy shell is going to be quickly exposed, finally leading to the contraction of the stellar radius during mass loss. For the He-MS stage, when the surface reaches the inside of the initial stellar envelope, the adiabatic mass loss would not lead to a rapid radius change due to an almost constant entropy profile. Such details of radiative response are introduced in Paper II \citep{PaperII}.
On the other hand, the convective envelope gives us another picture of radius evolution. A convective envelope has a less dense structure. During the adiabatic mass loss, the high-entropy area is exposed due to the superadiabatic zone. The helium stars expand rapidly and keep the maximum radius until the convective envelope is wholly ejected. Such details of superadiabatic expansion response are very similar to the behavior of the RGB star. It was introduced in Paper I \citep{PaperI}.

With the help of adiabatic mass loss response, we now consider the specific situation in helium binary mass transfer. 
A non-negligible He-HG stage issue is the delayed unstable mass transfer. The mass transfer rate is strongly related to the difference between stellar and Roche-lobe radius, so reaching the dynamical mass transfer rate at the beginning is impossible. The mass transfer rate will likely gradually reach a thermal timescale mass transfer rate. Then, internal energy will not be able to be rebalanced inside the stellar envelope, which will finally accelerate to an adiabatic mass transfer. 
Here, we use the method that we introduced in Paper I \citep{PaperI}, replacing the stellar radius with the radius inside the envelope, which could cause the thermal timescale mass transfer rate. 

We use Kelvin–Helmholtz timescale ($\tau_{\textrm{KH}}$) as a good approximation of thermal timescale and calculate the thermal timescale mass transfer rate as $\dot{M}_{\textrm{KH}} = M_{\textrm{He}} / \tau_{\textrm{KH}}$. In most cases, the binary mass transfer will be stable if $\dot{M}<\dot{M}_{\textrm{KH}}$. If the Roche-lobe radius ($R_\textrm{L}$) is equal to the radius $R_{\textrm{KH}}$, the mass transfer rate will equal to $\dot{M}_{\textrm{KH}}$. As a critical situation, the unstable mass transfer quickly appears once Roche-lobe reaches $R_{\textrm{KH}}$, then adiabatic mass loss shall happen in the binary system.

\begin{figure*}[ht!]
\plotone{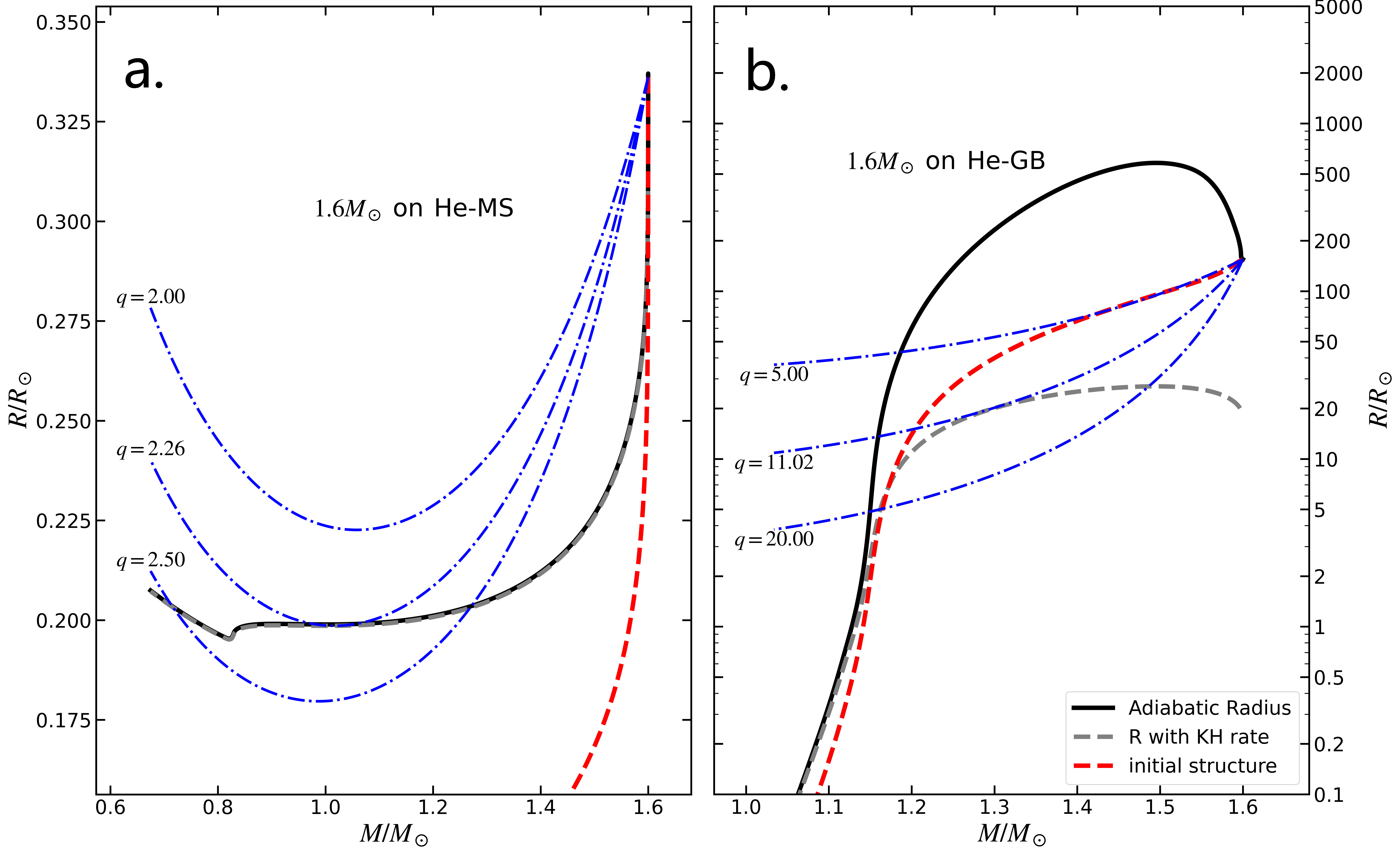}
\caption{Radius evolution tracks of two $1.6\,M_{\odot}$ helium stars during adiabatic mass loss within the mass coordinate. In sub-figure 'a', we show the total radiative structure of a He-TAMS star. In sub-figure 'b', it is a convective envelope on the giant branch. The black solid line is the radius change. The red dash line is the initial radius profile. Because of adiabatic expansion, we can see that the surface radius is always larger than the initial profile. Using the method as the trigger of unstable mass transfer in Paper I \citep{PaperI}, we show the maximum radius of Roche-lobe as a grey dash line. Assuming binary mass transfer is conservative and the mass ratio is known, the $R_\textrm{L}$ can be rebuilt as a blue dash line during mass transfer.
\label{fig:1.6adiab}}
\end{figure*}

Stellar radius and $R_{\textrm{KH}}$ responses are described in Figure \ref{fig:1.6adiab}. The black solid line represents the radius response, and the grey dashed line is $R_{\textrm{KH}}$. Clearly, the donor star can not start the unstable RLOF at the beginning of binary mass transfer in this sample. To show the boundary reaches the limit of $\dot{M}_{\textrm{KH}}$, we introduce a mass–radius exponent $\zeta_{\textrm{ad}}$,
\begin{equation}
\zeta_{\textrm{ad}}=\left ( \frac{\partial \ln_{}{R_{\textrm{KH}}}  }{\partial \ln_{}{M}}  \right ) _{\textrm{KH}} 
\end{equation} \\
It represents the donor star's adiabatic response to mass loss and we have already introduced it in Paper I \citep{PaperI}. Here, we choose a He-TAMS and a He-GB model to introduce how to get $q_{\textrm{crit}}$ in the helium binary system. We can easily conclude that $R_{\textrm{KH}}$ is much smaller than the stellar radius on He-GB, but He-TAMS does not show such scene instead. Actually, that is special for convective envelopes due to the low-density area. In this work, we assume a conservative situation for binary mass transfer, which means all mass through RLOF from helium donor transfer to accretor, and no material is lost from the binary system. Paper I \citep{PaperI} has shown how the law of conservation of mass and angular momentum would influence the $R_\textrm{L}$. In conservative mass transfer, the $R_\textrm{L}$ shall determined by the initial mass ratio ($q$). We define it as equal to donor mass divided by accretor mass (here, the donor is especially represented by a helium star).
\begin{equation}
q=M_{\textrm{He}}/M_{\textrm{acc}}
\end{equation}
Blue lines show the $R_\textrm{L}$ by different $q$. If it is constantly larger than $R_{\textrm{KH}}$, the mass transfer rate will not be high enough to start the adiabatic mass loss. Keep increasing $q$, unstable mass transfer stage comes out when $R_\textrm{L}$ is lower than $R_{\textrm{KH}}$. Here, we define the minimum initial mass ratio $q$ as the stability criteria $q_{\textrm{crit}}$, just letting $R_\textrm{L}$ tangents with $R_{\textrm{KH}}$.  

\begin{figure*}[ht!]
\plotone{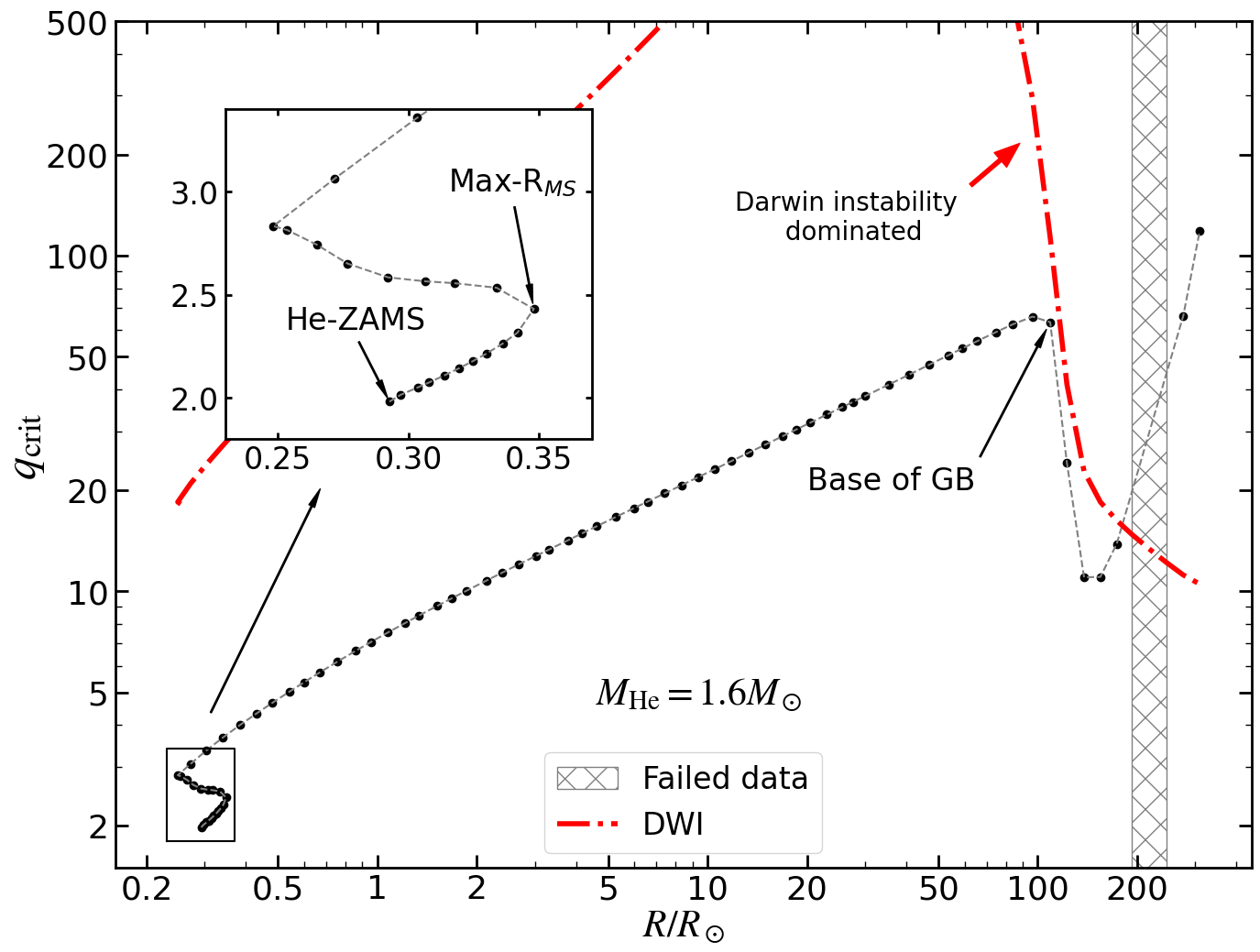}
\caption{$q_{\textrm{crit}}$ for $1.6\,M_{\odot}$ helium stars from He-ZAMS to C ignition. The results keep increasing on He-MS and He-HG. Due to the development of the convective envelope, $q_{\textrm{crit}}$ firstly decreases at the beginning of He-GB and then increases at later He-GB because of the radius increasing along with evolution. The 'Failed data' area is for the models that we can not appropriately simulate due to the steep entropy profile at the superadiabatic zone near the stellar surface. The red dash dot line is the maximum mass ratio of Darwin instability (DWI), which is discussed in Section \ref{subsec:discussion}.
\label{fig:1.6qcrit-total}}
\end{figure*}

Using the method, we calculated the $q_{\textrm{crit}}$ during $1.6\,M_{\odot}$ helium star evolution track using the selected models in Section \ref{sec:sequences}. The stability criteria are shown in Figure \ref{fig:1.6qcrit-total}. In this figure, $q_{\textrm{crit}}$ is a function of radius. With the evolution of the helium star, the radius is increasing on the early He-MS and He-HG/GB, as we introduced in Section \ref{sec:sequences}.
In the early He-MS and He-HG stage, due to radius increasing, $q_{\textrm{crit}}$ have to get larger to reach the $R_{\textrm{KH}}$ profile. Things are quite different in the later He-MS and He-GB. For the shrinking stage on later He-MS, the radius of the inner envelope changes faster than the surface (see the difference in Figure \ref{fig:1.6entropy-radius}-b), and $q_{\textrm{crit}}$ is still getting larger alone with stellar evolution. After the He-HG stage, the convective envelope quickly evolves inward from the surface. When the mass of the surface convective envelope increases to around $10^{-3}\,M_{\odot}\textrm{/yr}$, the stellar mass transfer suddenly becomes unstable due to the development of the low-density envelope. With a helium star ascending the He-GB, the convective zone dredges inward and the stellar radius increases rapidly. After the beginning of He-GB, the quickly increasing radius and the short thermal timescale dominate  the stability criteria, and $q_{\textrm{crit}}$ increases during most of He-GB.

For He-GB stars, our simulations face a great challenge in calculating adiabatic expansion ( the failed data is shown in Figure \ref{fig:1.6qcrit-total}). In this narrow parameter space, the radius is increasing too fast to calculate the appropriate response for adiabatic mass loss. This difficulty is caused by the increases of the entropy in the superadiabatic zone. The mass loss brings a higher entropy surface and leads to faster expansion. Due to the positive feedback, our simulation fails to solve the structure response in the superadiabatic zone around the surface area, which is similar to the situation on AGB stars in Paper III \citep{PaperIII}. 
When it comes to a later stage before carbon ignition, we notice that the superadiabatic zone evolves to a less steep entropy profile (see Figure \ref{fig:1.6adiab}). We calculated the stability criteria here. The structure of helium stars on He-GB is similar to low and intermediate-mass normal RGB stars (both a convective envelope with a degenerate and non-degenerate core). Comparing the $q_{\textrm{crit}}$ tendency on RGB in Paper II \citep{PaperII}, we have not seen special changes on the giant branch. So here we assume the $q_{\textrm{crit}}$ is increasing at a later stage of He-GB.

In this section, we introduced the method of calculating the stability criteria of adiabatic mass loss and the results of a $1.6\,M_{\odot}$ helium star. Generally, stellar models with higher $q_{\textrm{crit}}$ become more stable during adiabatic mass loss. It might not be reasonable for helium stars to have a much more stable mass transfer channel on He-HG/GB ($q_{\textrm{crit}}>10$). We expected it to be dominated by the short thermal timescale mass transfer during outer Lagrangian point ($L_3$) at these stages. It will be discussed in Section \ref{subsec:discussion}. 

\subsection{Critical Mass Ratio In M-R Parameter Spaces} 
\label{subsec:qcrit}

Using the same method in Subsection \ref{subsec:1.6 adiab}, we have calculated all the models in different mass sequences.

\begin{figure*}[ht!]
\plotone{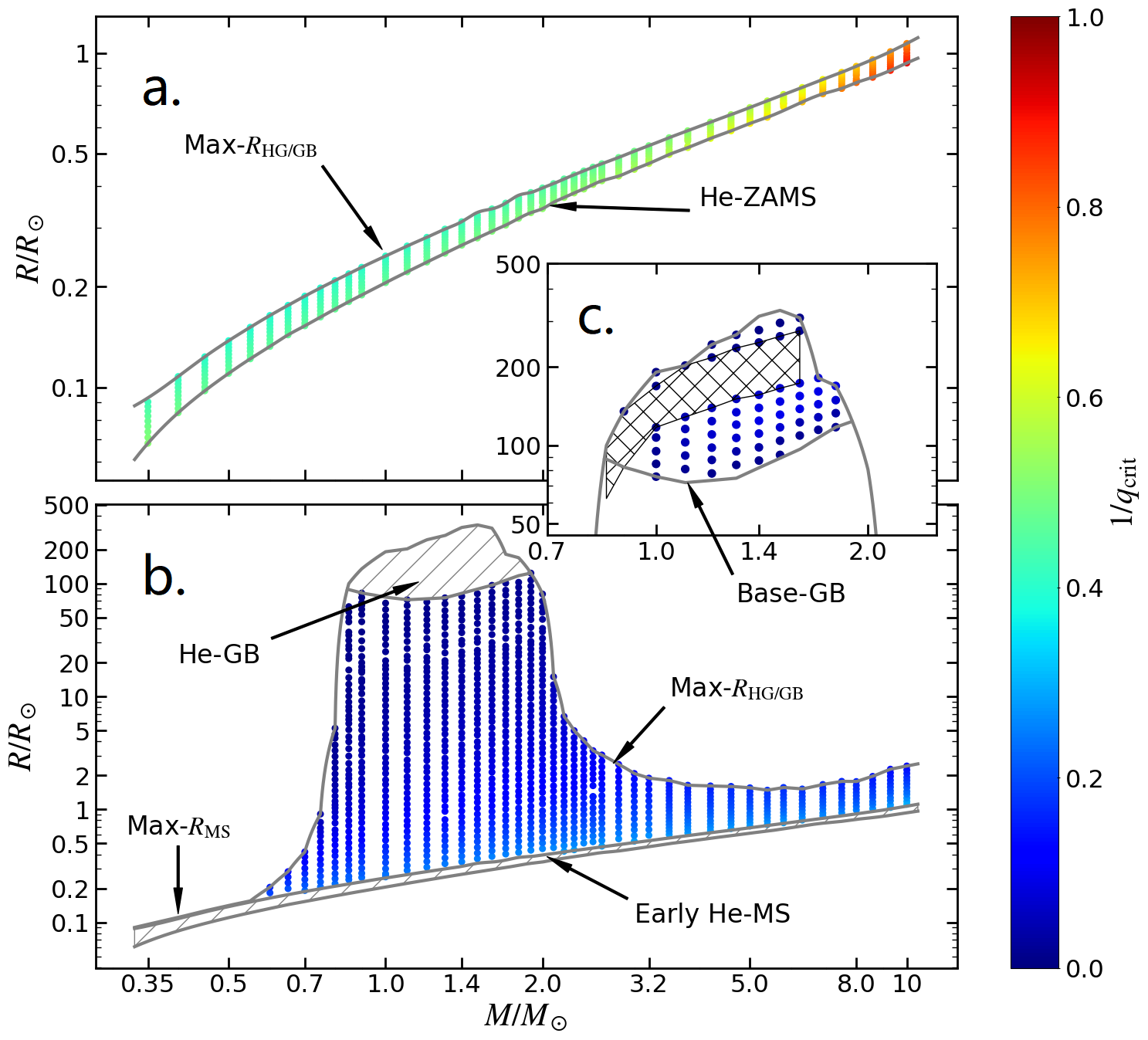}
\caption{Initial results of stability criteria for helium mass sequences in Mass-Radius ($M-R$) parameter space. We use $1/q_{\textrm{crit}}$ to highlight the lower results that $q_{\textrm{crit}}<10$. Here, we ignore the shrinkage at the later MS stage, and just consider the expanding phase of evolution. We show them in three parts: Early He-MS stage (Case BA) in sub-figure 'a'; He-HG stage (Case BBe) in sub-figure 'b'; and He-GB stage (Case BBl) in sub-figure 'c'. The shadow area represents failed data introduced in Section \ref{subsec:1.6 adiab}.
\label{fig:R-M_qcrit_initial}}
\end{figure*}

To show the key process of radius increasing, we use $M-R$ parameter space (also called Webbink diagram) to show the stability criteria of helium star mass sequences. Given an intuitive radius of the donor star, we could easily find the evolution stage at the beginning of the RLOF. 
There are several mechanisms to make the donor fill its Roche lobe. The most possible one is the radius expanding during evolution. Some processes (like magnetic braking (\citealt{1983ApJ...275..713R}), gravity wave, and tidal instability) could reduce the separation of the binary system to create the possibility for the donor to fill its Roche lobe even during the shrinkage stage. However, when the RLOF is undergoing, the angular momentum transfer with the mass flow will dominate the binary separation and period change. Other mechanism can be overridden at this stage.
It is obvious to find that stellar expansion is the primary stage for binary mass transfer. Suppose we ignore the possibilities of mass transfer on the shrinkage stage, dominated by quick binary orbit reducing, and just consider the donor radius expanding as the only reason for filling its Roche lobe. We can find all mass transfer phases on the $M-R$ diagram. Noticing that helium star evolves faster than the accretor, the initial mass of the helium star may be lower than the companion. In other words, the initial mass ratio of the mass transfer stage could have $q<1$. 

We define the Case BA phase as when helium stars fill their Roche-lobe at the He-MS stage. Case BA shall end at the maximum radius on He-MS, and the helium star could unlikely fill the Roche lobe in the later He-MS stage due to the decreasing radius. At the He-HG/GB stage, the helium star could fill the Roche lobe when the radius is larger than the maximum radius on He-MS, which we define as Case BB. Depending on the structure of the envelope, Case BB can be separated by Case BBe (radiative) in He-HG stage and Case BBl (convective) in He-GB stage.

\begin{figure*}[ht!]
\plotone{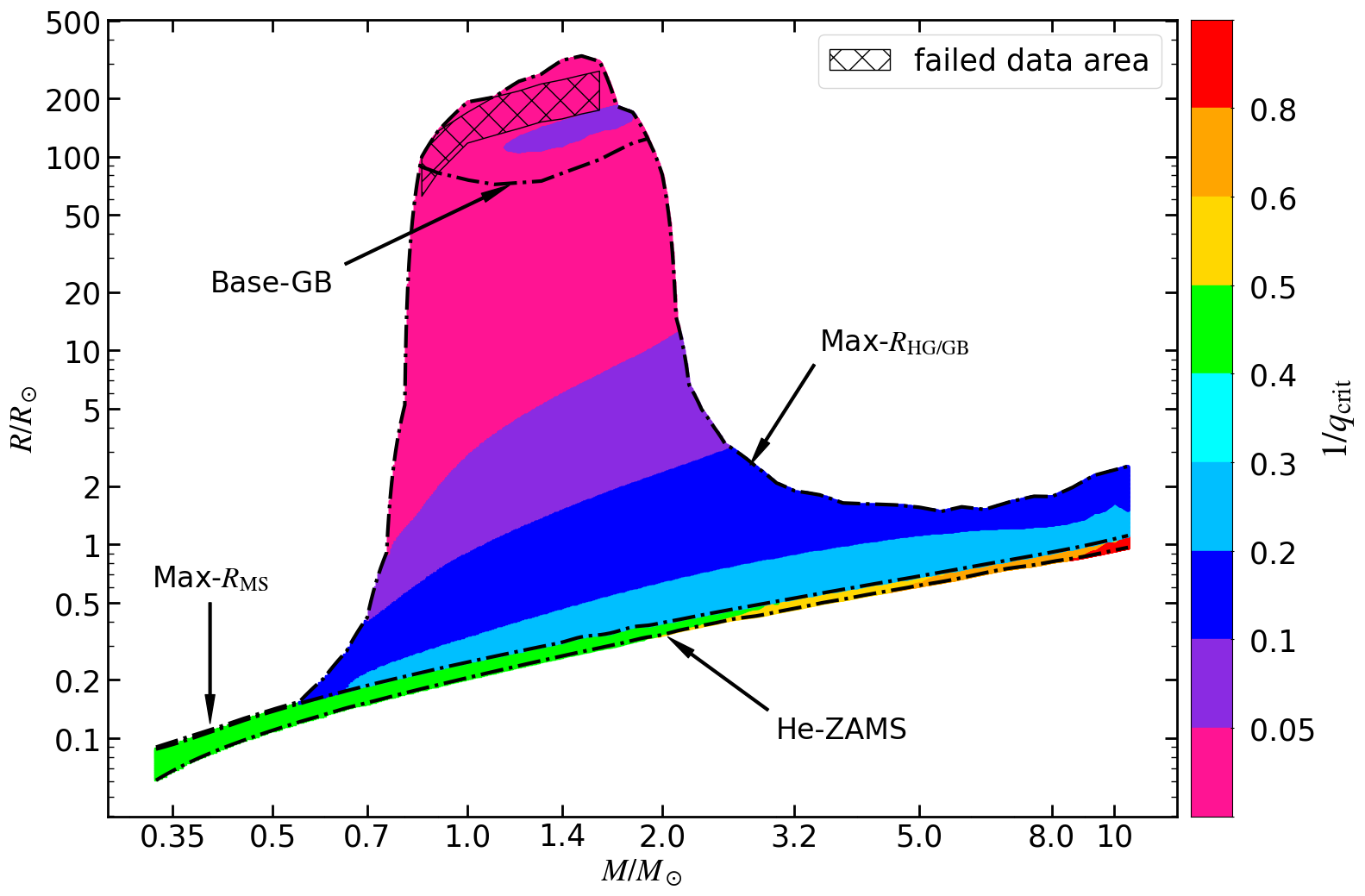}
\caption{Fitting $q_{\textrm{crit}}$ for helium mass sequences in $M-R$ parameter space. This result is using RBF interpolating to fit the initial result of $q_{\textrm{crit}}$. Shadow area represents failed data introduced in Subsection \ref{subsec:1.6 adiab}. To highlight the large results on He-HG/GB and following the observation tradition, we use $1/q_{\textrm{crit}}$ instead to show this parameter space.
\label{fig:R-M_qcrit}}
\end{figure*}

Figure \ref{fig:R-M_qcrit_initial} shows the $q_{\textrm{crit}}$ on the $M-R$ parameter space. In the early He-MS stage, the results of $q_{\textrm{crit}}$ are within the range of 1.0 to 2.6. On the He-HG stage, it ranges from 2.0 to several hundred. On early He-GB, adiabatic mass transfer quickly becomes unstable due to convective envelope, but $q_{\textrm{crit}}$ does not decrease below 10.0. 

The distribution of initial results can be used to fit the stability criteria on $M-R$ parameter space. Because of the continuity of stellar structure with mass increasing and evolution, the criteria $q_{\textrm{crit}}$ also shows continuity on the $M-R$ diagram. Due to the initial data being unevenly distributed, we use Radial Basis Function (RBF) interpolation to solve the fitting result. 
Figure \ref{fig:R-M_qcrit} shows fitting $q_{\textrm{crit}}$ for the helium sequences we studied. 

The fitting results are very suitable for binary population synthesis. Due to the extreme mass ratio being rare to be found and separated both in theory and observation, in Figure \ref{fig:R-M_qcrit}, we use $1/q_{\textrm{crit}}$ as the symbol instead of $q_{\textrm{crit}}$ and ignore the specific difference when $1/q_{\textrm{crit}}<0.05$. This way of expression is based on the observer's tradition. 

\section{Discussion} 
\label{subsec:discussion}
In the previous sections, we illustrated the method and concluded the results of the stability criteria in helium binary systems using the adiabatic mass loss model. In this part, we will analyze our results and compare them with other research and observations.

Our result show $1.0<q_{\textrm{crit}}<2.6$ for He-MS donor stars. During the evolution of a given mass helium star, $q_{\textrm{crit}}$ is getting larger from He-ZAMS to He-TAMS. However, with the increase of stellar mass,  $q_{\textrm{crit}}$ of the same evolutionary stage helium stars are getting smaller. This trend of helium stars is similar to normal MS stars. The criteria from He-HG to He-GB also have the same tendency as HG and RGB stars, but the critical mass ratios $q_{\textrm{crit}}$ are much larger (compared with Paper II by \citealt{PaperII}). This result is due to the much larger entropy profile of helium stars. Considering that unstable mass transfer should only occur when the mass ratio is larger than $q_{\textrm{crit}}$, our results may suggest that more binary systems tend to avoid dynamical timescale mass transfer for helium donor stars compared with ordinary HG and RGB donor stars.  

However, for some developed He-HG and 
He-GB stars, we notice that the $q_{\textrm{crit}}$ are extremely large and we barely see such mass ratio in observation. 
The extreme mass ratio systems may be dominated by the Darwin instability  (\citealt{1879RSPS...29..168D,2001ApJ...562.1012E,2019arXiv190701877S}), it will finally lead to the merging stage in a short timescale by the tidal effect. By assuming a tidally locked binary system and overriding the rotational angular momentum of secondary star, we can calculate the maximum mass ratio for helium binary system (\citealt{1995ApJ...444L..41R}).  We give the dimensionless gyration radius of helium star by using the moment of inertia. A sample is shown in Figure \ref{fig:1.6qcrit-total}. Luckily, compared with HG and GB stars, the helium stars are compact enough to bring a very small moment of inertia. As a result, Darwin instability (DWI) in the helium binary allows a very large critical mass ratio on He-HG/GB. After He-GB stage,The $q_{\textrm{crit}}$ in our research is larger than maximum mass ratio of DWI. Still, the actual rotation of helium and secondary star certainly decrease the maximum mass ratio. We can only give a rough prediction that DWI may influence our conclusion in the systems with high stellar rotation speed or after He-GB.

Furthermore, the stability criteria of helium stars on the He-GB stage are still uncertain. By the influence of a very short KH thermal timescale, the mass transfer could turn to the unstable phase even when $\dot{M} \leq \dot{M}_{\textrm{KH}}$ and the stellar radius will quickly reach the outer Lagrangian point. On the other hand, the thick shell of delayed unstable mass transfer during He-HG leads to thermal timescale mass transfer and causes unexpected expansion. 
These helium stars may undergo a thermal timescale mass transfer process \citep{2020ApJS..249....9G} before reaching the dynamical timescale mass transfer. In the future, we will study such processes in helium binary systems. 

This study calculates the critical mass ratio by assuming that mass and angular momentum are conserved during mass transfer. Fortunately, the non-conserved mass transfer will only change the orbital evolution. In other words, the donor's response during adiabatic mass loss is independent of the binary system's orbital evolution. We can easily present the critical mass ratio for non-conserved cases, such as the results applied in the double WDs study by \citet{2023A&A...669A..82L}. \citet{2024A&A...681A..31P} also provide a method to derive the critical mass ratio of non-conserved mass transfer based on Ge et al.'s series work, and they systematically study the forming of merging double compact objects.

\begin{figure*}[ht!]
\plotone{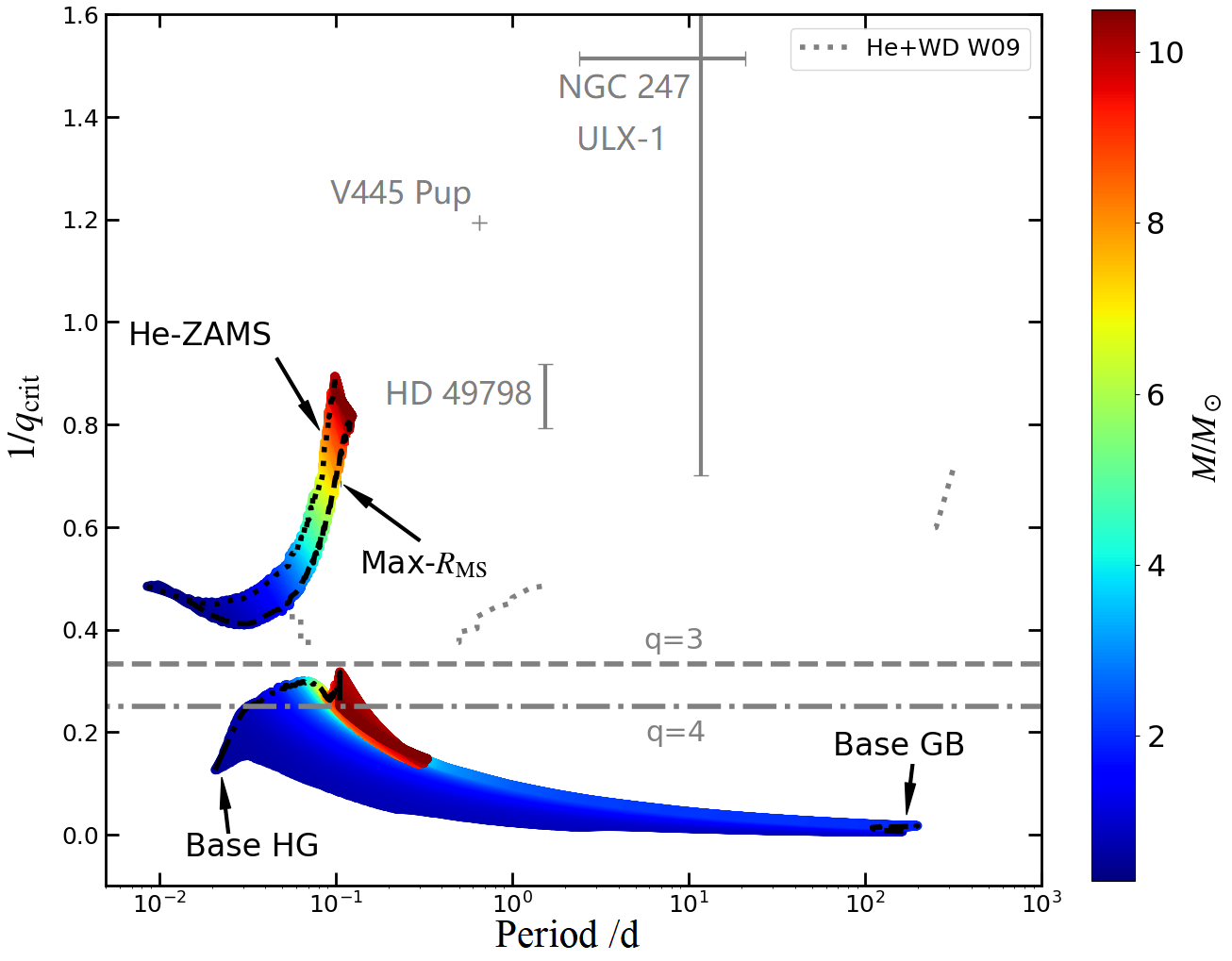}
\caption{The relation of $1/q_\textrm{crit}$ and stellar binary orbital period. The color bar represents the stellar mass. We light the color on the expansion stage of He-MS and He-HG. The dotted line is a critical boundary for the network of progenitors of SN Ia when $M_\textrm{accretor}=1.2$ by \citealt{2009MNRAS.395..847W}. Horizontal lines are the criteria of the polytropic models on He-MS and He-HG by \citealt{2014A&A...563A..83C}. The data K24 of $\phi$ Per-type binaries is from \citealt{2024ApJ...962...70K}. We also put two indirect helium binaries, NGC 247 ULX-1 and V445 Pup, in this figure.
\label{fig:P-q1}}
\end{figure*}

\begin{figure*}[ht!]
\plotone{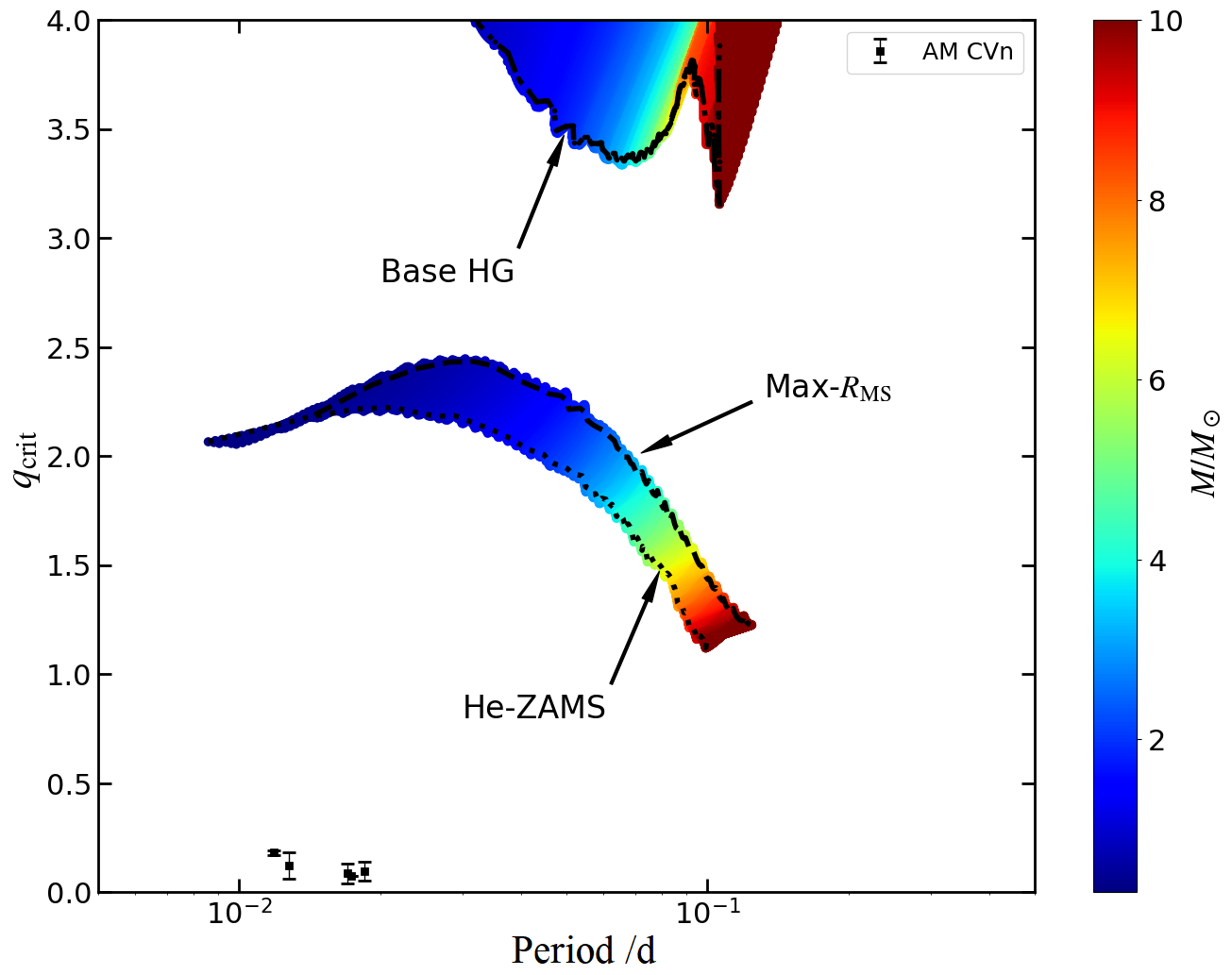}
\caption{Same as the previous figure, but use $q_\textrm{crit}$ instead. This figure focuses on the He-MS and early He-HG stage at the short-period side. The data points are the potential helium star channel AM CVn stars in \citealt{2010PASP..122.1133S}.
\label{fig:P-q2}}
\end{figure*}

Although we have given $q_{\textrm{crit}}$ in $M-R$ parameter space, the mass and radius can not be directly observed. The calculation of stellar mass and radius must introduce the stellar structure model, which causes unnecessary effects when compared with the stability criteria. 
Here, if we consider the stellar radius as the Roche lobe radius $R=R_\textrm{L}$, the relation of orbital period, mass, radius, and mass ratio is easily found (see Paper II by \citealt{PaperII}).
\begin{equation}
\begin{split}
\log{(P/\mathrm{d})} =&\frac{3}{2} \log{ (R_\mathrm{L}/R_\odot) } - \frac{1}{2} \log{ (M/M_\odot) } \\
&+ \log{ g(q) } - 0.45423,
\end{split}
\end{equation} \\
where $P$ represents the orbital period in days, $g(q)$ is a very weak function of mass ratio $q$ and we use $q_\textrm{crit}$ as the initial mass ratio:
\begin{equation}
g(q) =\left ( \frac{2q}{1+q}  \right )^\frac{1}{2} \left ( \frac{ 0.6+q^{-\frac{2}{3}}\ln{ (1+q^{\frac{1}{3}}) }  }{0.6+\ln{2} }  \right )^\frac{3}{2}.
\end{equation} \\
Using this relation, we change the criteria to the mass ratio-orbital period ($q-P$) parameter space and compare the critical results with other theories and observations. 
We recalculate the relation of $q_{\textrm{crit}}$ and binary orbital period in Figure\,\ref{fig:P-q1} and Figure\,\ref{fig:P-q2}. The color bar in these two figures represents the stellar mass. We light the mass on the expansion stage of He-MS and He-HG. 

In the polytropic model, the results of $q_{\textrm{crit}}$ are constant values. For He-MS stars, $q_{\textrm{crit}}=3$, for He-HG stars $q_{\textrm{crit}}=4$ and for He-GB stars $q_{\textrm{crit}}=0.78$ when the accretor is non-degenerate star (\citealt{2002MNRAS.329..897H,2014A&A...563A..83C}). Compared with the constant criteria of the polytropic model, the $q_{\textrm{crit}}$ in our research is evolving with the stellar status. In Figure \ref{fig:P-q1}, we give the specific difference. 
Our result indicates a more unstable parameter space compared with polytropic results of $q=3$ for He-MS donors. Our results suggest that more helium binary systems can evolve to unstable mass transfer phases like contact binaries, and fewer systems could evolve through stable mass transfer to become AM CVn or UCXB systems.
For He-HG donor stars, our results on parameter space cover both sides of $q=4$, which means the specific difference from polytropic model relies on stellar mass and structure. Generally speaking, it is more unstable than the polytropic model for early He-HG stars when $M_\textrm{He}>2$, but it is more stable on the other part which covers more parameter space. Thus, we predict the birth rate of SNe Ia should be more significant in the Case BB channel. More long period helium binaries will go through the stable mass transfer stage and produce long period double WD systems. However, further study of the realistic thermal timescale mass transfer still needs to confirm this prediction.

We also compare the network of helium binary progenitors of SN Ia (\citealt{2009MNRAS.395..847W,2014ApJ...794L..28W}). In their simulation, the mass transfer rate $\dot{M}>10^{-5}\,M_{\odot}\textrm{/yr}$ is considered as the criterion of the unstable mass transfer situation (see dot line in Figure\,\ref{fig:P-q1}). They lock the initial accretor's mass and change the periods and the initial helium star's mass (in this figure, we show the results when $M_\textrm{acc}=1.2$), which makes it difficult to compare. We noticed that the network is incomplete, which causes the stability criterion to be incoherent for some period.  Still, we can see our result on He-MS is more unstable, which can be explained by the bias of conservative mass transfer. The He-HG stage gives an opposite prediction. Firstly, their limitation may be too strict to cover the thermal timescale mass transfer into a stable situation. Secondly, delayed unstable mass transfer plays a significant role during the He-HG stage. 

For low-mass helium stars, there is an overlap between the possible mass ranges of helium stars and He WDs from $0.35\,M_{\odot}$ to $0.45\,M_{\odot}$. Sometimes, the observation cannot separate them clearly, especially in short orbital period systems.
We notice that the stability studies of NS with He WD systems have shown no space for unstable mass transfer on Mass-Period parameter space \citep{2022ApJ...930..134C}. Though the mass range covers $0.17\,M_{\odot}$ to $0.45\,M_{\odot}$, the structure difference between He WD to helium star is significant. By using our criteria, we expect the mass transfer to be stable if the donor star is a helium star at a similar mass range from $0.35\,M_{\odot}$ to $0.45\,M_{\odot}$ with the $1.4\,M_{\odot}$ NS companion.

As we seek the comparison object from observed helium star binaries, we find it challenging to test our result directly. The critical problem is the lack of helium binary systems at the early stage of mass transfer, though the possibility of this stage is high. If a helium star fills its Roche lobe on He-MS, the orbital period will be only a few minutes to hours. The possible binary objects will be AM CVns (accreting WD companion) and ultra-compact X-ray binaries (UCXBs; accreting NS companion), which are hydrogen-poor in the spectrum with degenerate companions. Due to the weak observational features, these systems make solving the orbital parameters and type of component difficult. Only a few AM CVns are believed to have gone through the helium channel and just passed the stable mass transfer (\citealt{2010PASP..122.1133S}). We plot them in Figure\,\ref{fig:P-q2}. However, for these systems, the mass of the donor is almost entirely lost. They have entered the late stage of mass transfer, which makes them less suitable to compare with the stable criteria. Besides, due to the $q_\textrm{crit}$ on He-MS is lower than the polytropic model, our results prevent more massive He-MS stars to become the donor of AM CVn systems. The upper mass limit is $M_\textrm{He}=0.95\,M_{\odot}$  when $M_\textrm{WD}=0.4\,M_{\odot}$ and $M_\textrm{He}=1.77\,M_{\odot}$  when $M_\textrm{WD}=0.8\,M_{\odot}$.

Our results are unsuitable to compare with some sdB/O binary systems, which are going through the last stage of stable mass transfer to strip the hydrogen envelope and expose its helium core. These systems are considered as stripped helium stars. Typical systems including VFTS 291 (\citealt{2023MNRAS.525.5121V}) and LB-1 (\citealt{2019Natur.575..618L,2020A&A...639L...6S}) do not start RLOF because of the evolution of helium star. In fact, they are likely preparing to reach He-ZAMS after the hydrogen shell is ejected. Hence, we do not analyze these systems in this article.

For mass-transferring helium stars on the He-HG/GB stage, the initial orbital periods are from days to hundreds of days. With the extremely short timescale of He-HG/GB stars and the limited mass range around $2.0\,M_{\odot}>M_{\textrm{He}}>0.9\,M_{\odot}$, We expect their possibility to be observed is slim. They will likely be found as low-mass X-ray binaries (LMXBs) with an NS companion. If the companion is a B-type MS, the accretor may spin up and become Be star due to angular momentum transfer. It may form the $\phi$ Per-type binary, which consists of sdO and Be star. Here, we use the recent data (\citealt{2024ApJ...962...70K}). The periods of these systems are longer than 60 days. Due to the massive mass of Be stars, the mass ratios of these systems are not near to our stability criteria. Still, they are not good candidates for comparison due to the mass ratio being around 0.05 to 0.2, which is far from our criteria for He-HG. However, the criteria suggest a stable mass transfer stage fitting the observation. Similar to observed AM CVns, they are not at the beginning of mass transfer.
In addition to the direct observation, some systems show indirect evidence of helium binary mass transfer. If the secondary is NS or BH, the stable mass transfer at a high accreting rate will create a high X-ray luminosity and become an ultra-luminous X-ray source (ULX). A potential ULX which has a helium donor is NGC 247 ULX-1. However, the distance is too far to detect the binary parameters. Recently, \citet{2023ApJ...947...52Z} gives a prediction of helium star mass by assuming a $1.4\,M_\odot$ NS accretor ($M_\textrm{He}=0.6\sim2.0\,M_{\odot}$, $P_\textrm{orb}=2.4\sim21\, $d), and the rough result is shown on Figure\,\ref{fig:P-q1}.
Besides, the only known helium nova V445 Pup is believed to have evolved from a WD accreting mass from a helium donor due to the dominant helium lines in the nova's spectrum. A helium nova outburst caused by the helium flash of the accreting shell on the WD surface is the most favorable explanation for this event, which reveals a stable mass transfer stage before this event. Here, we use the result ($M_\textrm{WD}=0.8\,M_{\odot}$, $M_\textrm{He}=0.67\,M_{\odot}$ by using $d=8.2\,$kpc) calculated by \citet{2023ApJ...952L..26B}. It is also shown in Figure\,\ref{fig:P-q1}. We can easily find that neither is a good limit for stability criteria. All the systems are far from the $q_\textrm{crit}$.

For some sdB/O binaries, they have yet to reach the RLOF stage. Using our criteria, we could predict the mass transfer stage. We take the eclipsing sdO binary system HD49798 as a sample ($M_\textrm{WD}=1.28\pm0.05\,M_{\odot}$, $M_\textrm{He}=1.50\pm0.05\,M_{\odot}$ and $P=1.55\,$d by \citealt{2017ApJ...847...78B}). The mass ratio and period are shown in Figure\,\ref{fig:P-q1}. Due to the $q<q_\textrm{crit}$ in the $q-P$ parameter space, the Case BB mass transfer should be stable. This method can also be used to guide helium binary population synthesis.

As we can see, there have been few detections of helium binaries during RLOF up to now. The observation of helium binaries is still strongly required. We hope there will be more detections, especially for high-time resolution surveys, to find more systems to compare with different criteria. These systems will be very important for us to understand the mass-transfer evolution of helium stars.

\section{Summary} \label{summary}

This study attempts to give stability criteria of dynamical mass transfer in low and intermediate-mass helium binary systems. In this article, we use the adiabatic mass loss model (Paper I \citealt{PaperI}, II \citealt{PaperII}, III \citealt{PaperIII}) to study the stellar structure response of helium donor during binary mass transfer. 

We first simulate the evolution of a single helium star as the donor. To get the most developed structure of He-HG/GB stars, we ignore the stellar wind of helium stars. In our simulation, $M_{\textrm{He}}<0.6\,M_{\odot}$ helium stars can only go through the He-MS mass transfer stage. Helium stars in $2.0\,M_{\odot}>M_{\textrm{He}}>0.9\,M_{\odot}$ could experience the fierce expansion and create a convective envelope on the He-GB. For $M_{\textrm{He}}>2.0\,M_{\odot}$ donors, the convective envelope fails to be built by the premature carbon ignition. 

Then, we analyze the adiabatic mass loss response of different helium donor stars. After we get the radius change of the adiabatic mass loss process, we rebuild the Roche-lobe radius and compare them in the different mass ratio systems. We conclude the results by giving the stability criteria $q_\textrm{crit}$ on $M-R$ diagram. Here, the binary mass and angular momentum transfer are conserved. Generally speaking, the $q_{\textrm{crit}}$ increases with the evolution of the helium star. For main-sequence helium donor stars, the results show $1.0<q_{\textrm{crit}}<2.6$. After early He-HG, the $q_{\textrm{crit}}$ quickly increases larger than 10 for helium donors. With increased stellar mass, the $q_{\textrm{crit}}$ of the He-MS is getting smaller. This result is similar to the tendency of a typical star from MS to RGB.

In the last section, we compare the results with different stability criteria. Our result indicates a more unstable parameter space for He-MS donors than the polytropic model, which indicates that more He-MS shall go through CE or contact binary systems, and the birth rate of SNe Ia through Case BA should be lower. Our critical mass ratios for the early He-HG donors are smaller than previous results, which is more unstable. However, the $q_\textrm{crit}$ rapidly increases after early He-HG and becomes highly stable. Such extreme mass ratio systems may only be influenced by Darwin instability on late He-GB. The specific evolution at this stage may rely on further research on the non-conserved thermal timescale mass transfer.

Finally, we compare our results with some observed helium binary systems going through RLOF. For AM CVns and $\phi$ Per-type binaries, our criteria do fit these objects, but the mass ratio of these systems varies far from $q_\textrm{crit}$. Other indirect helium binaries like helium nova and sdB ULXs are also not good candidates for validating our theoretical results. Our results can be applied to the binary population synthesis code to study helium binary systems.

\section{acknowledgments}
\label{ack}
This project is supported by the National Natural Science Foundation of China (NSFC, grant Nos. 12288102, 12090040/3, 12173081), National Key R\&D Program of China (grant Nos. 2021YFA1600403, 2021YFA1600401), the Key Research Program of Frontier Sciences, CAS (No. ZDBS-LY-7005), CAS-Light of West China Program, Yunnan Fundamental Research Projects (Nos. 202101AV070001, 202201BC070003), Yunnan Revitalization Talent Support Program - Science \& Technology Champion Project (No. 202305AB350003), and International Centre of Supernovae, Yunnan Key Laboratory (No. 202302AN360001). L.Z thanks the selfless help of H.G, X.C and Z.H. L.Z also appreciate Jingxiao.Luo and Luhan.Li for the guidance of possible observation objects of helium binaries.

\appendix



\section{Physical parameters and mass transfer stability criteria of helium stars}
Here we show the data of mass sequence $1.6\,M_{\odot}$ as samples. The stellar radius versus $q_{\textrm{crit}}$ in this table is shown in Figure\,\ref{fig:1.6qcrit-total}. The completed data of mass sequences are released in the machine readable table. The columns of Table\,\ref{tab:test} are as follows:\\
1. k --- sequence model number, indicate the order from He-ZAMS to C ignition;\\
2. age --- evolution age measured from He-ZAMS model (k=1 model);\\
3. $M_\textrm{CO}$ --- carbon/oxygen core mass in solar unit;\\
4. log $R$ --- initial helium stellar radius before mass loss;\\
5. log $t_\textrm{{KH}}$ --- Kelvin–Helmholtz timescale of the initial model;\\
6. log $L$ --- initial helium stellar luminosity before mass loss;\\
7. log $T_{\rm eff}$ --- initial helium stellar effective temperature before mass loss;\\
8. log $g$ --- the acceleration of gravity on the stellar surface;\\
9. $k^2$ --- the dimensionless gyration radius. $k^2=I/MR^2$, $I$ is moment of inertia, $M$ is stellar mass, $R$ is stellar radius;\\
10. $\Psi_c$ --- the degeneracy at the center of helium star;\\
11. $X_{\textrm{He}}$ --- helium fraction at the center;\\
12. $\zeta_{\textrm{ad}}$ --- mass–radius exponent at $\dot{M}=\dot{M}_{\textrm{KH}}$. The '...' marker represents the failed data;\\
13. $q_{\textrm{crit}}$ --- critical mass ratio. The '...' marker represents the failed data;\\
14. type --- evolutionary stage of helium star;\\

\begin{table}[htbp]
    \centering
    \caption{Physical parameters and mass transfer stability of a $1.6\,M_{\odot}$ helium star} 
    \label{tab:test}
    \begin{tabular}{cccccccccccllc}
        \hline
        k & age & $M_\textrm{CO}$ & log $R$ & log $t_\textrm{{KH}}$ & log $L$ & log $T_{\rm eff}$ & log $g$ & $k^2$ & $\Psi_c$ & $X_{\textrm{He}}$ & $\zeta_{\textrm{ad}}$ & $q_{\textrm{crit}}$ & type \\ 
        {} & yr & $M_{\odot}$ & $R_{\odot}$ & yr & $L_{\odot}$ & ${\rm K}$ & ${\rm cm/s^2}$ & {} \\
        \hline
        01  &  0.00e+00  &  0.62  &  -1.2282  &  5.3630  &  3.0737  &  1.5680  &  5.7092  &  0.076423  &  -2.178  &  0.96232  &   2.567  &   1.981 & \multirow{16}{*}{He-MS} \\ 
        02  &  7.93e+05  &  0.65  &  -1.2134  &  5.3355  &  3.0948  &  1.5684  &  5.6964  &  0.074497  &  -2.183  &  0.85903  &   2.639  &   2.015  \\ 
        03  &  1.30e+06  &  0.68  &  -1.1914  &  5.3055  &  3.1152  &  1.5684  &  5.6772  &  0.072549  &  -2.199  &  0.76800  &   2.717  &   2.051  \\ 
        04  &  1.62e+06  &  0.70  &  -1.1777  &  5.2864  &  3.1284  &  1.5685  &  5.6653  &  0.071325  &  -2.209  &  0.71338  &   2.767  &   2.075  \\ 
        05  &  2.08e+06  &  0.73  &  -1.1586  &  5.2589  &  3.1476  &  1.5687  &  5.6488  &  0.069580  &  -2.223  &  0.63831  &   2.842  &   2.110  \\ 
        06  &  2.51e+06  &  0.75  &  -1.1412  &  5.2328  &  3.1661  &  1.5688  &  5.6336  &  0.067936  &  -2.235  &  0.57052  &   2.915  &   2.144  \\ 
        07  &  2.91e+06  &  0.78  &  -1.1252  &  5.2078  &  3.1842  &  1.5690  &  5.6197  &  0.066380  &  -2.246  &  0.50849  &   2.987  &   2.177  \\ 
        08  &  3.36e+06  &  0.80  &  -1.1085  &  5.1803  &  3.2045  &  1.5693  &  5.6052  &  0.064672  &  -2.257  &  0.44245  &   3.070  &   2.216  \\ 
        09  &  3.89e+06  &  0.83  &  -1.0901  &  5.1473  &  3.2295  &  1.5698  &  5.5893  &  0.062643  &  -2.267  &  0.36666  &   3.173  &   2.264  \\ 
        10  &  4.42e+06  &  0.85  &  -1.0739  &  5.1140  &  3.2558  &  1.5705  &  5.5751  &  0.060584  &  -2.273  &  0.29236  &   3.283  &   2.316  \\ 
        11  &  5.53e+06  &  8.85  &  -1.0556  &  5.0474  &  3.3144  &  1.5727  &  5.5593  &  0.052898  &  -2.255  &  0.14475  &   3.530  &   2.431  \\ 
        12  &  6.57e+06  &  0.92  &  -1.0975  &  5.0059  &  3.3740  &  1.5776  &  5.5956  &  0.052897  &  -2.153  &  0.03628  &   3.750  &   2.534  \\ 
        13  &  6.83e+06  &  0.93  &  -1.1472  &  5.0070  &  3.3945  &  1.5809  &  5.6388  &  0.052143  &  -2.064  &  0.01369  &   3.798  &   2.556  \\ 
        14  &  6.91e+06  &  0.93  &  -1.1827  &  5.0114  &  3.4055  &  1.5831  &  5.6697  &  0.051834  &  -2.001  &  0.00672  &   3.816  &   2.565  \\ 
        15  &  6.96e+06  &  0.93  &  -1.2300  &  5.0146  &  3.4229  &  1.5861  &  5.7107  &  0.051228  &  -1.907  &  0.00221  &   3.857  &   2.584  \\ 
        16  &  6.98e+06  &  0.93  &  -1.2845  &  5.0007  &  3.4605  &  1.5904  &  5.7581  &  0.049233  &  -1.740  &  0.00014  &   4.001  &   2.652  \\

        \hline
        17  &  6.98e+06  &  0.93  &  -1.3273  &  4.9769  &  3.5029  &  1.5945  &  5.7953  &  0.046641  &  -1.473  &  0  &   4.194  &   2.742 & \multirow{20}{*}{He-HG} \\ 
        20  &  7.01e+06  &  0.93  &  -1.3027  &  4.8610  &  3.6081  &  1.5987  &  5.7739  &  0.037867  &  -0.803  &  0  &   4.878  &   3.062  \\ 
        23  &  7.04e+06  &  0.94  &  -0.9596  &  4.5189  &  3.8012  &  1.5934  &  5.4759  &  0.024070  &  -0.278  &  0  &   6.860  &   3.995  \\ 
        26  &  7.08e+06  &  0.96  &  -0.6154  &  4.2417  &  3.9289  &  1.5846  &  5.1769  &  0.016051  &   0.348  &  0  &   9.032  &   5.020  \\ 
        29  &  7.11e+06  &  0.98  &  -0.2796  &  4.0070  &  4.0177  &  1.5742  &  4.8852  &  0.011134  &   0.923  &  0  &  11.467  &   6.173  \\ 
        32  &  7.13e+06  &  0.99  &  0.0713  &  3.7898  &  4.0826  &  1.5617  &  4.5804  &  0.007811  &   1.459  &  0  &  14.336  &   7.535 \\ 
        35  &  7.14e+06  &  1.00  &  0.4144  &  3.5969  &  4.1264  &  1.5483  &  4.2824  &  0.005597  &   1.943  &  0  &  17.495  &   9.040 \\ 
        38  &  7.15e+06  &  1.01  &  0.7598  &  3.4161  &  4.1573  &  1.5339  &  3.9824  &  0.003997  &   2.385  &  0  &  21.111  &  10.768 \\ 
        41  &  7.15e+06  &  1.01  &  1.1045  &  3.2447  &  4.1790  &  1.5188  &  3.6831  &  0.002837  &   2.775  &  0  &  25.240  &  12.745 \\ 
        44  &  7.15e+06  &  1.02  &  1.4282  &  3.0887  &  4.1944  &  1.5041  &  3.4019  &  0.002043  &   3.091  &  0  &  29.720  &  14.897 \\ 
        47  &  7.16e+06  &  1.02  &  1.7907  &  2.9182  &  4.2075  &  1.4872  &  3.0870  &  0.001401  &   3.415  &  0  &  35.571  &  17.713 \\ 
        50  &  7.16e+06  &  1.02  &  2.1209  &  2.7662  &  4.2160  &  1.4714  &  2.8002  &  0.000985  &   3.675  &  0  &  41.810  &  20.723 \\ 
        53  &  7.16e+06  &  1.02  &  2.4709  &  2.6077  &  4.2225  &  1.4541  &  2.4962  &  0.000673  &   3.914  &  0  &  49.569  &  24.475 \\ 
        56  &  7.16e+06  &  1.03  &  2.8246  &  2.4494  &  4.2272  &  1.4363  &  2.1890  &  0.000459  &   4.117  &  0  &  58.886  &  28.990 \\ 
        59  &  7.16e+06  &  1.03  &  3.1355  &  2.3115  &  4.2301  &  1.4203  &  1.9189  &  0.000337  &   4.269  &  0  &  68.506  &  33.662 \\ 
        62  &  7.16e+06  &  1.03  &  3.4013  &  2.1945  &  4.2317  &  1.4063  &  1.6880  &  0.000280  &   4.388  &  0  &  77.840  &  38.202 \\ 
        65  &  7.16e+06  &  1.03  &  3.8480  &  1.9993  &  4.2329  &  1.3824  &  1.3000  &  0.000276  &   4.550  &  0  &  96.621  &  47.356 \\ 
        68  &  7.17e+06  &  1.03  &  4.1820  &  1.8543  &  4.2328  &  1.3640  &  1.0100  &  0.000420  &   4.650  &  0  &  113.639  &  55.667  \\ 
        71  &  7.17e+06  &  1.03  &  4.5686  &  1.6878  &  4.2314  &  1.3422  &  0.6741  &  0.001966  &   4.849  &  0  &  134.536  &  65.890  \\

        \hline
        72  &  7.17e+06  &  1.03  &  4.6919  &  1.6364  &  4.2292  &  1.3350  &  0.5670  &  0.005537  &   5.272  &  0  &  129.514  &  63.432   & \multirow{10}{*}{He-GB} \\ 
        73  &  7.17e+06  &  1.03  &  4.8085  &  1.5768  &  4.2383  &  1.3290  &  0.4658  &  0.017204  &   6.235  &  0  &  48.879  &  24.141   \\ 
        74  &  7.17e+06  &  1.04  &  4.9281  &  1.4616  &  4.3015  &  1.3263  &  0.3619  &  0.034318  &   8.458  &  0  &  21.630  &  11.016   \\ 
        75  &  7.18e+06  &  1.04  &  5.0420  &  1.3476  &  4.3660  &  1.3240  &  0.2629  &  0.043957  &  11.863  &  0  &  21.696  &  11.048   \\ 
        76  &  7.18e+06  &  1.05  &  5.1558  &  1.2345  &  4.4297  &  1.3216  &  0.1641  &  0.050977  &  16.669  &  0  &  27.539  &  13.849   \\ 
        77  &  7.18e+06  &  1.05  &  5.2599  &  1.1315  &  4.4875  &  1.3195  &  0.0737  &  0.057009  &  22.408  &  0  &  ...  &  ...  \\
        78  &  7.19e+06  &  1.05  &  5.3876  &  1.0061  &  4.5575  &  1.3167  &  -0.0372  &  0.064529  &  31.089  &  0  &  ...  &  ...  \\
        79  &  7.19e+06  &  1.06  &  5.5021  &  0.8939  &  4.6199  &  1.3142  &  -0.1366  &  0.071669  &  40.711  &  0  &  ...  &  ...  \\
        80  &  7.19e+06  &  1.07  &  5.6200  &  0.7766  &  4.6860  &  1.3118  &  -0.2391  &  0.079331  &  54.250  &  0  &  135.291  &  66.260    \\ 
        81  &  7.20e+06  &  1.11  &  5.7341  &  0.6566  &  4.7564  &  1.3099  &  -0.3382  &  0.085401  &  74.684  &  0  &  241.300  &  118.316   \\ 
        \hline 
    \end{tabular}
    \tablecomments{Table 1 is published in its entirety in the machine-readable format.\\
      A portion is shown here for guidance regarding its form and content.}
\end{table}

\bibliography{Adiab-4}{}
\bibliographystyle{Adiab-4}



\end{document}